\documentclass{revtex4-2}
\usepackage{graphicx}
\usepackage{physics}
\usepackage{xcolor}
\usepackage{amsmath}
\usepackage{mathtools}
\usepackage{stmaryrd}
\usepackage{longtable}[=v4.13]

\makeatletter
\g@addto@macro\@mkpream@relax{\let\@endpbox\relax}
\makeatother
\usepackage{amssymb}
\usepackage{comment}
\usepackage[left=1in, right=1in]{geometry}

\usepackage[normalem]{ulem}

\DeclareMathOperator{\sgn}{sgn}
\DeclareMathOperator{\Var}{Var}
\DeclareMathOperator{\Spec}{Spec}
\DeclareMathOperator{\diag}{diag}

\newcommand{\jj}{\jmath}

\begin{document}

\title{{\bf Breaking of clustering and macroscopic coherence under the lens of asymmetry measures}}
\author{Florent Ferro}
\affiliation{Universit\'e Paris-Saclay, CNRS, LPTMS, 91405, Orsay, France}

\begin{abstract}
    In one-dimensional systems, spontaneous symmetry breaking (SSB) states are fragile by nature,
    as the injection of a non-zero energy density above the ground state is expected to restore
    the symmetry. This instability implies that local perturbations can lead to macroscopic
    correlation profiles, a breaking of clustering properties and even macroscopic quantum
    superpositions. In this work, we investigate the effect of interaction on this phenomenology
    by considering an interacting model with conserved domain wall number, that possesses a
    ferromagnetic ground state breaking the $\mathbb Z_2$ symmetry of the Hamiltonian. We first
    show that a local quench in this system amplifies quantum interferences, producing a
    macroscopic magnetisation profile that directly reflects the scattering phase of the model.
    Then, we use two asymmetry measures, namely the Entanglement Asymmetry (EA) and Quantum Fisher
    Information (QFI), to characterise the quantum coherence associated with the fluctuations of
    the magnetisation. By focusing on subsystems comparable in size to the light-cone of the
    perturbation, we confirm the emergence of macroscopic quantum coherence throughout the whole
    perturbed region. Finally, we discuss the link between EA and QFI and show that the
    variance/EA inequality for pure state can be generalised to a QFI/EA inequality for mixed
    states.
\end{abstract}

\maketitle

\section{Introduction}
While quantum physics is very efficient to describe the smallest scales of matter, it is still
unclear whether genuine quantum phenomena can survive at macroscopic scales. For equilibrium
states of systems described by local Hamiltonians, clustering properties ensure that at large
distances, different subsystems act independently, leaving no ground for macroscopic quantum
superpositions such as the famous ``cat states" -or GHZ states \cite{Greenberger1989} - imagined
by Schrödinger \cite{schrodinger1935gegenwartige, trimmer1980present}. The interest in macroscopic
quantum effects however came back with the recent breakthroughs in quantum simulators and quantum
circuits, in which it is now possible to engineer such superpositions of up to $\sim 20$ qubits
\cite{Gao2010Experimental, omran2019generation, song2019generation}. However, the protocols to
create such states are fine tuned, and by nature cat states are fragile and unstable: the loss of
a single qubit destroys the superposition. Therefore, two questions arise: can macroscopic quantum
superpositions emerge from more ``natural" scenarios, and if so, can they be more robust than cat
states?

In equilibrium, critical ground states are natural candidates since they possess algebraically
decaying correlations, allowing for ``enhanced" quantum coherence
\cite{DiFresco2022Multiparameter, DiFresco2024Metrology, chen2026quantum}.
Out-of-equilibrium, it is natural to directly look for scenarios in which correlations are
non-decaying altogether in a given region. This breaking of clustering is typically signalled by
magnetization profiles becoming self-similar at different times through a rescaling (ballistic,
diffusive...). Such scaling dynamics occur in various settings: local quenches on ferromagnetic
ground states \cite{Zauner2015Time, Eisler2020Front, Delfino2022Space} or quantum scars
\cite{Bocini2024Growing}, domain wall melting \cite{Antal1999Transport, Scopa2023Scaling,
Eisler2025Domain}, systems with jamming \cite{Bidzhiev2022Macroscopic, Fagotti2024Quantum},
interface physics \cite{Hunyadi2004Dynamic, Collura2018Analytic, Eisler2013Full,
maric2025disorder, maric2025macroscopic}... In particular, in Ref
\cite{ferro2025kickingquantumfisherinformation} we showed that ``robust" macroscopic quantum
superpositions can appear after a local quench on the spontaneously symmetry broken (SSB) ground
state of the Ising model in the ordered phase. This effect has a very simple interpretation:
excitations in the symmetry-broken phase are not local but semi-local, which essentially means
that they are semi-classically described by domain walls \cite{Sachdev1997LowTemperature,
Rieger2011Semiclassical}. The ballistic spreading of a domain wall wavefunction results in quantum
superposition of states with macroscopically different (diverging linearly in time)
magnetisations. Such a state is thus similar to a kink version of a W state \cite{Dur2000Three},
which is much more robust than cat states in the sense that it survives the loss of a part of the
system (note that W states can also be prepared experimentally, see for instance Refs
\cite{Haffner2005Scalable, catalano2025experimental}).

It is not obvious however whether this phenomenology can survive in presence of interactions,
which induces non-trivial scattering and the formation of bound states. In this paper, we fill
this gap and generalise the results of 
\cite{ferro2025kickingquantumfisherinformation} on local quenches to the interacting case. We
study quantum fluctuations by explicitly computing two quantities from the resource theory of
asymmetry \cite{korzekwa2013resource}, having in mind that an asymmetry with respect to the
symmetry group generated by a charge $Q$ essentially quantifies coherence between the charge
sectors of $Q$ \cite{Marvian2016How, Marvian2016Quantum, Yamaguchi2023Beyond, Zhang2017Detecting}.

The first of these tools is the Entanglement asymmetry (EA), also known as G asymmetry
\cite{gour2024resourcesquantumworld} or relative entropy of superposition
\cite{aberg2006quantifyingsuperposition} in the field of quantum information. Since its original
introduction for the study of the quantum Mpemba effect \cite{Ares2023Entanglement,
ares2025quantummpembaeffects}, EA has recently gained a lot of attention in the many-body and
high-energy communities \cite{Murciano2024Entanglement, Chalas2024Multiple,
Yamashika2024Entanglement, Yu2025Quantum, Teza2026Speedups, Yu2025Tuning, Liu2025Symmetry,
turkeshi2024quantum, banerjee2025entanglement, Ares2025Quantum, Caceffo2024Entangled,
Rylands2024Dynamical, Klobas2024NonEquilibrium, Khor2024Confinement, Bertini2024Dynamics,
Ferro_2024, russotto2025non,Benini2025Entanglement,
chen2025entanglementasymmetryhaydenpreskillprotocol, Kusuki2025Entanglement,
Fossati2024Entanglement, capizzi2023entanglementasymmetryorderedphase,
AresPiroli2025Entanglement}, and has even been measured experimentally \cite{Joshi2024Observing,
xu2025observation}. It is now recognised as a standard tool to quantify how much a quantum state
breaks a symmetry. By bounding the asymmetry of subsystems of size comparable to the light-cone,
we will extract its leading behaviour and show that subsystems fall in quantum superposition of a
number of magnetisation sectors that diverges in time.

To complete this analysis, we also compute the Quantum Fisher Information (QFI), and explicitly
work out a relation between them through a new bound based on the one between entropy and
variance. While the QFI can be used to quantify asymmetry \cite{gour2024resourcesquantumworld,
takagi2019Skew, kusuki2026resource, yamashika2025quantum}, it was originally introduced in the
context of quantum metrology via the Cramer Rao bound: the QFI determines the maximal precision
one can achieve in parameter estimation. Actually, Ref~\cite{Tan2021Fisher} showed that QFI is a
probe for resourceful quantum states in a large number of resource theories. One of its main
interests relies in the fact that it gives a lower bound to the entanglement depth (the minimum
number of multipartite entangled parts of a given state), making it a witness of multipartite
entanglement \cite{Pezze2009Entanglement, Hyllus2012Fisher, Toth2012Multipartite,
Froewis2012Measures}. Furthermore, contrary to the entanglement depth, QFI is not sensitive to infinitesimally
small correlations between different parts of the system. For these reasons, it is a sensible
probe of multipartite entanglement in many-body physics, and has been used to characterise it both
in \cite{Frerot2016Quantum, Lambert2019Estimates, chen2026quantum} and out of equilibrium
\cite{ferro2025kickingquantumfisherinformation, Zhong2014Quantum, Paviglianiti2023multipartite}, including in monitored quantum circuits~\cite{Lira2025multipartite}. While the QFI is generally a
very complex quantity to evaluate, a lot of techniques have been developed to access it, notably
in free-fermionic systems \cite{Jiang2014Quantum, Gao2014Bounds, safranek2015quantum}, and through
simpler related quantities such as the Wigner-Yanase-Dyson Skew Information
\cite{Wigner1963INFORMATION}, see ref \cite{ferro2025kickingquantumfisherinformation}.
Furthermore, it can be expressed as a series of experimentally accessible quantities
\cite{Rath2021Quantum}, and in thermal states its connection to the susceptibility
\cite{Frerot2016Quantum} makes it directly measurable \cite{Hauke2016Measuring}. 

We use the dual XXZ spin chain \cite{Zadnik2021Dual} --- obtained from the XXZ chain via a
Kramers-Wanier transformation --- as an interacting toy model to study the effect of a local
perturbation on an SSB ground state that spontaneously breaks the $\mathbb{Z}_2$ symmetry of the
Hamiltonian. We take advantage of its Coordinate Bethe Ansatz description to solve the dynamics
after the quench, and show that, as expected, it breaks the clustering properties on ballistic
scales through its transverse magnetisation profile. By ``diagonalising" the scattering phase of
the model, we make analytical predictions for the scaling limit of the EA and QFI that we compare
with exact numerical data, obtained using the low-dimensionality of the accessible Hilbert space
after the local quench. We show that interaction does not spoil the phenomenology occurring in
free models, and that any subsystem close to the perturbation, with a size scaling with that of
the light-cone, falls into robust macroscopic quantum superpositions. The main results are summarised here in Table~\ref{tab:results}.

\setlength{\LTcapwidth}{%
    \dimexpr 0.92\linewidth+4\tabcolsep\relax
}
\begin{longtable}{p{0.22\linewidth}
    @{\hspace{1cm}}
    p{0.65\linewidth}}
    
    \label{tab:results}\\
    \hline
    Observable & Main result \\
    \hline
    \endfirsthead

    \hline
    Observable & Main result \\
    \hline
    \endhead

    Magnetisation/ correlations
    &
    Macroscopic magnetisation profile and breaking of clustering on
    ballistic scales. Quantum interferences make the scattering
    phase directly accessible through the magnetisation profile.
    Bound states do not contribute to the breaking of clustering.
    \\[6pt]

    Entanglement asymmetry (EA)
    &
    Logarithmic growth in time associated with coherence between an
    increasing number of magnetisation sectors. Its leading
    coefficient is determined by scattering states, while
    bound states contribute only a bounded term.
    \\[6pt]

    Quantum Fisher information (QFI)
    &
    Quadratic scaling with subsystem size at intermediate times, confirming macroscopic quantum coherence
    in the interacting model. Several bounds on the QFI are derived to approximate it at any desired level of precision.
    \\[6pt]

    Relation between EA and QFI
    &
    A lower bound on QFI in terms of EA, extending the
    pure-state entropy--variance relation to mixed states.
    Sufficiently large logarithmic EA implies super-extensive QFI.
    \\

    \hline
    \\
    \caption{
    Summary of the main results.
    The out-of-equilibrium results concern a local spin flip over the
    ferromagnetic product state $\ket \Uparrow$, at large times and subsystem sizes.
    Macroscopic coherence after local quenches was studied
    in free models in Ref.~\cite{ferro2025kickingquantumfisherinformation};
    here we investigate its persistence in the presence of
    interaction.
    The EA--QFI bound is independent of this quench protocol.
    }
\end{longtable}

\section{Setup}\label{sec:setup}
In this paper, we consider the dual XXZ Hamiltonian \cite{Zadnik2021Dual}:
\begin{equation}\label{eq:hamiltonian}
    H = \frac 1 4 \sum_{j=1}^N(\Delta -\sigma^x_j)(1-\sigma^z_{j-1}\sigma^z_{j+1})
    -h\sum_{j=1}^N\sigma_j^z \sigma^z_{j+1},
\end{equation}
where we choose to impose periodic boundary conditions (PBC) $\sigma^\alpha_{N+1} =
\sigma^\alpha_1$. For simplicity, since $(\prod_{i=1}^N \sigma^z_i) H(\Delta, h)
(\prod_{i=1}^N \sigma^z_i) = -H(-\Delta, -h)$ we will furthermore assume $\Delta\ge 0$.
By duality with the XXZ model, this Hamiltonian is exactly solvable by Coordinate Bethe Ansatz
\cite{Franchini2017An}. It presents a $U(1)$ symmetry generated by the charge $Q = \frac 1 2
\sum_{j=1}^N(1-\sigma^z_j \sigma^z_{j+1})$, which counts the number of domain walls between
ferromagnetic regions of opposite transverse magnetisations. Additionally, the system exhibits a
$\mathbb{Z}_2$ symmetry represented by the spin flip operator $P = \prod_{j=1}^N \sigma^x_j$. In
the zero particle sector ($Q=0$), there are two eigenstates of $H$ corresponding to each
eigenvalue of $P$:
\begin{equation}
    \ket{\pm} = \frac{1}{\sqrt 2}(\ket{\uparrow}^{\otimes N} \pm \ket{\downarrow}^{\otimes N}),
\end{equation}
such that $P \ket{\pm} = \pm \ket \pm$. We can anticipate that for sufficiently high values of
$h$, ferromagnetic interactions dominate and the ground state of $H$ lies in the $Q=0$ sector. It
is easy to see that $\ket +$ and $\ket -$ are degenerate, $H\ket{\pm} = -hN \ket{\pm}$, therefore the trivial
product states with maximal and minimal transverse magnetisation, $\ket \Uparrow =
\ket{\uparrow}^{\otimes N}$ and $\ket \Downarrow = \ket{\downarrow}^{\otimes N}$, are also
eigenstates of $H$. These states explicitly break the $\mathbb{Z}_2$ symmetry of the system.

We are interested in the effect of a local perturbation on the symmetry-broken ground state, $\ket
\Uparrow$. The simplest such perturbation is the creation of a small ferromagnetic domain of
neighbouring sites $\llbracket j_1, j_2\rrbracket = \{j_1, j_1+1, \cdots, j_2\}$, of size $d_0 =
j_2 - j_1 + 1$. Without loss of generality ($H$ is translationally invariant) and assuming $d_0$
is odd for simplicity, we center this domain around site $0$ ($\equiv N$ by PBC). We can thus
write the state after the local perturbation as:
\begin{equation}\label{eq:psi_t}
\ket{\Psi_{d_0}(t)} = e^{-iHt} \prod_{j= -(d_0-1)/2}^{(d_0-1)/2}\sigma^x_j \ket{\Uparrow }.
\end{equation}
Since $[H, Q]=0$ the state remains in the $Q=2$ sector at all times.
The field term in the Hamiltonian is $-hN + 2h Q$. Therefore, it only produces a global phase in
this sector and can be set to $0$ for simplicity when studying the dynamics. The field is however
relevant for the state preparation, since it ensures that the initial ferromagnetic state is a
ground state. In particular, the critical ground state at $h=0$ and $\Delta<1$ is not the initial
state of this protocol, which does not undergo a phase transition across $\Delta=1$. 

This means that the two domain walls can move and the distance between them can fluctuate, but
they cannot be created or annihilated. The physical situation corresponding to this local quench
is therefore similar to the one studied in Refs \cite{ferro2025kickingquantumfisherinformation,
Eisler2020Front}, except that the domain walls are not freely propagating but can instead scatter
or form bound states.

\section{CBA description}\label{sec:CBA}
We can explicitly work out the time evolution of the state given by Eq.~\eqref{eq:psi_t} using
coordinate Bethe Ansatz. Let us index the domain wall positions with half integers $l_1, l_2\in
\llbracket \frac 1 2, N-\frac{1}{2}\rrbracket$. For $l_1<l_2$ we define the basis states of the
$Q=2$ sector as:
\begin{equation}
    \ket{l_1 l_2}_{\pm} = \frac{1}{\sqrt 2}(\ket{\cdots\uparrow\uparrow \downarrow_{l_1+1/2}\cdots
    \downarrow_{l_2-1/2}\uparrow\uparrow \cdots}\pm \ket{\cdots\downarrow\downarrow
    \uparrow_{l_1+1/2}\cdots \uparrow_{l_2-1/2}\downarrow\downarrow \cdots}),
\end{equation}
where $P\ket{l_1 l_2}_\pm = \pm \ket{l_1 l_2}_\pm$. The boundary conditions depend on the parity
sector, $\ket{l_1 N+\frac 1 2}_\pm = \pm \ket{\frac 1 2 l_1}$, and also allow us to write negative
domain wall positions. In this basis the state $\ket{\Psi_{d_0}(t)}$ is expressed as a
superposition of states in different parity sectors:
\begin{equation}
    \ket{\Psi_{d_0}(t)} = \frac{1}{\sqrt 2}e^{-iHt} \left(\ket{-\frac{d_0}{2}, \frac{d_0}{2}}_+
    +\ket{-\frac{d_0}{2}, \frac{d_0}{2}}_-\right).
\end{equation} 

We show in Appendix~\ref{ap:thermo_amplitudes} that in the thermodynamic limit $N\to \infty$, for
fixed time $t$ and distances between the position of the domain walls, the transition amplitude
$\prescript{}{s}{\bra{f_1 f_2}}e^{-iHt} \ket{i_1 i_2}_s$ becomes independent of the parity sector $s$ and $N$. The
periodic chain is effectively ``unwrapped" into an infinite line of origin $0$. The indices $i_1,
i_2, f_1, f_2$ remain at a finite distance from the origin, and the paths with non-zero winding
number around the chain are suppressed by Lieb-Robinson bounds \cite{Lieb1972Finite}. 
\begin{figure}[t]
    \centering
    \includegraphics[width=\linewidth]{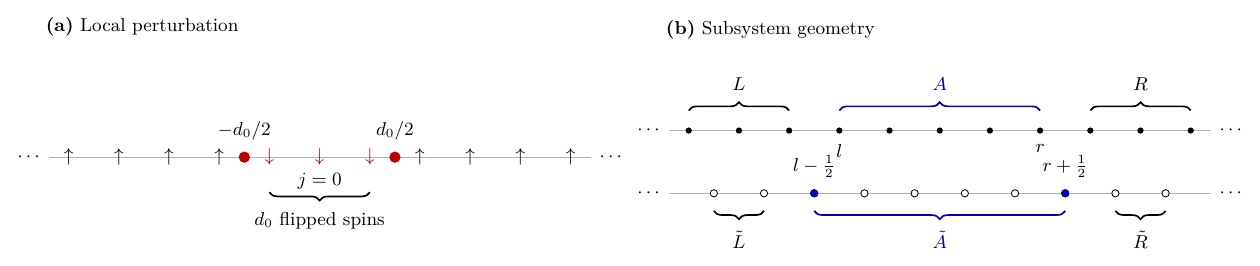}
    \caption{%
    Local quench and subsystem geometry after taking the
    thermodynamic limit.
    (a) A domain of $d_0$ neighbouring spins is flipped over the
    fully up state, creating two domain walls at
    $-d_0/2$ and $d_0/2$ ($d_0=3$ above).
    (b) A subsystem $A=\llbracket l,r\rrbracket$ and its left and
    right complements $L$ and $R$.
    Black dots represent integer spin sites, while empty
    circles represent half-integer domain wall positions.
    The corresponding regions are $\tilde L$, $\tilde A$ and
    $\tilde R$. The two boundary bonds, highlighted in blue,
    belong to $\tilde A$ by convention.}
    \label{fig:protocol}
\end{figure}

The local quench on the infinite chain is illustrated in
Fig.~\ref{fig:protocol}. For later use, we also introduce a subsystem
$A=\llbracket l,r\rrbracket$, its left complement
$L=\llbracket-\infty,l-1\rrbracket$ and its right complement
$R=\llbracket r+1,+\infty\rrbracket$. The corresponding regions of
domain wall positions are
$\tilde A=\llbracket l-\frac12,r+\frac12\rrbracket$,
$\tilde L=\llbracket-\infty,l-\frac32\rrbracket$ and
$\tilde R=\llbracket r+\frac32,+\infty\rrbracket$.

Consequently, it makes sense to introduce the states $\ket{l_1 l_2} = \frac{1}{\sqrt 2} (\ket{l_1
l_2}_+ + \ket{l_1 l_2}_-)$ that, after taking the thermodynamic limit, satisfy 
\begin{equation}
    \sigma^z_\jj\ket{l_1 l_2} = \sgn(l_1-\jj)\sgn(l_2-\jj)\ket{l_1 l_2}.
\end{equation}
This allows us to write the state at time $t$ as
\begin{equation}\label{eq:psi_evol}
    \ket{\Psi_{d_0}(t)} = \sum_{f_1<f_2\in \mathbb Z +\frac 1 2} \Psi_{f_1, f_2}^{d_0}(t) \ket{f_1 f_2}.
\end{equation}

The dynamics of the amplitude $\Psi_{f_1, f_2}^{d_0}(t)$ is governed by two types of excitations:
scattering states and bound states. We decompose the total wavefunction as a sum of these two
contributions: $\Psi_{f_1, f_2}^{d_0}(t) = \psi^{d_0}_{f_1, f_2}(t)+\nu^{d_0}_{f_1, f_2}(t)$ .
From now on, we subtract the constant energy \(2\Delta+4h\) in the two-domain-wall sector, which amounts to an irrelevant global phase of the full state.
\\
\paragraph{}
The first contribution $\psi^{d_0}_{f_1, f_2}(t)$ corresponds to pairs of particles with real
momenta $k_1, k_2\in \mathbb{R}$:
\begin{equation}\label{eq:magnons}
\psi^{d_0}_{f_1, f_2}(t) = \frac{1}{4\pi^2}\int_{-\pi}^\pi dk_1 dk_2\, \tilde \psi_{d_0}(k_1, k_2)
e^{i[k_1 (f_1-d_0/2) + k_2 (f_2+d_0/2)]}e^{-i(\epsilon_{k_1}+\epsilon_{k_2})t},
\end{equation}
where $\epsilon_k = -\cos k$ is the single-particle bare energy, $\tilde \psi_{d_0}(k_1, k_2) =
e^{id_0(k_1-k_2)} + S(k_1, k_2)$ and
\begin{equation}
    S(k_1, k_2) = -\frac{1+e^{i(k_1+k_2)}-2\Delta e^{i k_1}}{1+e^{i(k_1+k_2)}-2\Delta e^{i k_2}}
\end{equation}
is the scattering phase of the model. These excitations represent
scattering states (analogous to magnons in the XXZ model).
The stationary points of the phase in
Eq.~\eqref{eq:magnons} satisfy the equations
$\epsilon'_{k_{1,2}}=\sin k_{1,2}=f_{1,2}/t$. Real stationary points therefore exist for
$|f_1|,|f_2|\le t$. According to the standard stationary-phase
expansion, they give the leading contribution to the integral in the limit of large $t$. 
More precisely, the standard stationary-phase
expansion gives a factor $t^{-1/2}$ for each momentum integral,
and thus a contribution $t^{-1/2}\times t^{-1/2}=t^{-1}$
from each stationary point.
Outside this region there are no real stationary points,
and the integral is suppressed by the highly oscillatory integrand.
The stationary-point conditions thus describe two domain walls
spreading over a region of size $2t$.

Note that for $\Delta<1$, this integral expression is ill-defined, but it still makes sense by
representing the scattering phase in Fourier space, $S(k_1, k_2) = \sum_{n_1 n_2}S_{n_1 n_2}e^{i(k_1
n_1+k_2 n_2)}$, see Appendix \ref{ap:S_diag}. Under this representation, we can write
$\psi^{d_0}_{f_1, f_2}(t)$ as a sum over Bessel functions $J_n(t)$ as
\begin{equation}
    \psi^{d_0}_{f_1, f_2}(t) =i^{f_1+f_2}J_{f_1+\frac{d_0}{2}}(t)J_{f_2-\frac{d_0}{2}}(t)+
    \sum_{mn}i^{m+n}S_{m-f_1\frac{d_0}{2}, n-f_2-\frac{d_0}{2}}J_{m}(t)J_{n}(t).
\end{equation}
The numerical data presented below are obtained by evaluating
the full wavefunction, including both scattering and bound-state
contributions. For the scattering contribution, we use the
Bessel-function representation above and truncate the sums
to discard Bessel functions in their exponentially decaying
region. The bound-state contribution is evaluated from
Eq.~\eqref{eq:bound_states}. This calculation is performed
directly in the thermodynamic limit and does not involve
exact diagonalisation of a finite chain.

\paragraph{}
The other contribution $\nu^{d_0}_{f_1, f_2}(t)$ represents bound states. They correspond to pairs
of complex-conjugated momenta $\{K/2-i\eta(K), K/2+i\eta(K)\}$ with total momentum $K$, where
$e^{-\eta(K)} = \Delta^{-1}\cos (K/2)$ is the solution of $S(K/2+i\eta(K), K/2-i\eta(K)) = 0$.
Their contribution to the amplitude reads:
\begin{equation}\label{eq:bound_states}
    \nu_{f_1,f_2}^{d_0}(t) = \frac{1}{2\pi}\int_{-\pi}^{\pi} dK\, \tilde \nu_{f_2-f_1}^{d_0}(K)
    e^{iK\frac{f_1+f_2}{2}}e^{-iE_Kt},
\end{equation}
where
\begin{equation}
    \tilde \nu_{d_f}^{d_0}(K) = (e^{-\eta(K)(d_0 + d_f -2)}-e^{-\eta(K)(d_0 + d_f
    )})\theta(\eta(K)),
\end{equation}
$E_K = \epsilon_{K/2-i\eta(K)}+\epsilon_{K/2+i\eta(K)} = -\Delta-\Delta^{-1} \cos^2 (K/2)$ and
where the Heaviside step function $\theta(\eta(K))$ restricts the integration to states satisfying
$|\cos K/2| \le \Delta$. Therefore, while every momentum is allowed for $\Delta>1$, it
is not the case anymore when $\Delta<1$. This condition ensures that the bound-states
wavefunctions are normalisable and that their group velocity, given by $E_K' =
\frac{1}{2\Delta}\sin(K)$, does not exceed that of the scattering states (i.e. $1$).
$\nu_{f_1,f_2}^{d_0}(t)$ is $\mathcal O(t^{-1/2})$ and is exponentially decaying with $f_2-f_1$,
therefore it describes a localised ferromagnetic domain of $\mathcal{O}(1)$ size moving through
the chain in the light-cone $(-v_{B}^{\text{max}}t, v_{B}^{\text{max}} t)$, where the maximal
bound state velocity is given by $v_{B}^{\text{max}} = 1/(2\Delta)$ for $\Delta\ge 1/\sqrt 2$, and $v_{B}^{\text{max}} = \sqrt{1-\Delta^2}$ for $\Delta \le 1/\sqrt 2$.

\paragraph*{}
Using the wavefunctions \eqref{eq:magnons} and \eqref{eq:bound_states}, we introduce the
(unnormalised) states $\ket{\psi_{d_0}(t)} = \sum_{f_1<f_2\in \mathbb Z}\psi^{d_0}_{f_1 f_2}(t)
\ket{f_1 f_2}$ and $\ket{\nu_{d_0}(t)} = \sum_{f_1<f_2\in \mathbb Z}\nu^{d_0}_{f_1 f_2}(t)
\ket{f_1 f_2}$. These states correspond to the projection of $\ket{\Psi_{d_0}(t)}$ onto the
sectors associated with pairs of real and complex momenta respectively. $H$ is block diagonal
between these two sectors, therefore the norms of these states are constant in time. They are given
by
\begin{equation}\label{eq:proba_scattering}
    p^S_{d_0} = \braket{\psi_{d_0}(t)}{\psi_{d_0}(t)} = \frac{1}{8\pi^2}\int_{-\pi}^{\pi}dk_1 dk_2
    |\tilde \psi_{d_0}(k_1, k_2)|^2, 
\end{equation}
and $p^B_{d_0} = 1-p^S_{d_0}$.

\section{Scaling limit of the transverse magnetisation}\label{sec:scaling}
In this section, we study the scaling limit of the one and two-point correlation functions of the
transverse magnetisation after a spin flip ($d_0=1$), $\mathcal M_t(\jj) =
\bra{\Psi(t)}\sigma^z_\jj \ket{\Psi(t)}$ and $\mathcal M_t(\jj, \jj') = \bra{\Psi(t)}\sigma^z_\jj
\sigma^z_{\jj'} \ket{\Psi(t)}$. The basis $\ket{l_1 l_2}$ appearing in the decomposition
\eqref{eq:psi_evol} of the state is particularly convenient, as it is an eigenbasis for the local
transverse magnetisation operators,
\begin{equation}\label{eq:mag_basis}
    \sigma^z_\jmath\ket{l_1 l_2} = \sgn(l_1-\jmath)\sgn(l_2-\jmath)\ket{l_1 l_2}.
\end{equation}
Said otherwise, in the state $\ket{l_1 l_2}$ the magnetisation is $-1$ between the two domain
walls, and $+1$ otherwise. Since spins are associated to integer positions and domain walls to
half-integer ones, there is no ambiguity in the above expression. Using that $\sgn x =
2\theta(x)-1$, and inserting Eq.~\eqref{eq:mag_basis} into the expression of the correlation
functions, we can express both quantities in terms of the same operator:
\begin{equation}
    \left|
    \begin{aligned}
       \mathcal M_t(\jj) &= 1 -2 \bra{\Psi(t)} \Pi_{\llbracket -\infty,
       \jmath-1/2\rrbracket}^{\llbracket \jmath + 1/2, +\infty \rrbracket}\ket{\Psi(t)}, \\
        \mathcal M_t(\jj, \jj') &= 1-2\bra{\Psi(t)}\Pi_{\llbracket \jj+1/2,
        \jj'-1/2\rrbracket}^{\llbracket \jj'+1/2, +\infty\rrbracket}+\Pi_{\llbracket -\infty,
        \jj-1/2\rrbracket}^{\llbracket \jj+1/2, \jj'-1/2\rrbracket}\ket{\Psi(t)},
    \end{aligned}
    \right.
\end{equation}
where
\begin{equation}
    \Pi_{X}^Y = \sum_{\substack{l_1\in X, l_2\in Y\\l_1 < l_2}} \ket{l_1 l_2}\bra{l_1 l_2}.
\end{equation}

Physically, for two sets of neighbouring half-integers $X = \llbracket x_l, x_r\rrbracket$ and $Y
= \llbracket y_l, y_r\rrbracket$ the expectation value of the projector,
\begin{equation}\label{eq:defproba}
    p_{(X, Y)}(t) =\bra{\Psi(t)} \Pi_{X}^{Y}\ket{\Psi(t)},
\end{equation}
gives the probability to find one domain wall in $X$ and
the other in $Y$ at time $t$. We now proceed to define its scaling limit: we take $|x_l|, |x_r|,
|y_l|, |y_r| \to \infty$ while keeping the ratios $x_{l, r}/t$ and $y_{l, r}/t$ (the rays) fixed,
and maintaining a constant initial domain size $d_0=1$:
\begin{equation}\label{eq:scaling_pXY}
    \underline p_{(\mathcal X, \mathcal Y)} = \lim_{\substack{|x_{l, r}|, |y_{l, r}|, t\to \infty
    \\ (x_{l}/t, x_r/t) \to \mathcal X \\ (y_{l}/t, y_r/t) \to \mathcal Y}}p_{(X, Y)}(t).
\end{equation}
In this expression, the discrete intervals $X = \llbracket x_l, x_r\rrbracket$ are associated to
continuous ray intervals $\mathcal X = (x_l/t, x_r/t)$. As shown in appendix \ref{ap:scaling}, we
can decompose the scaling limit \eqref{eq:scaling_pXY} into a contribution coming from the
scattering states $\bra{\psi(t)} \Pi_{X}^{Y} \ket{\psi(t)}$ and one from the bound states
$\bra{\nu(t)} \Pi_{X}^{Y} \ket{\nu(t)}$ as $\underline p_{(\mathcal X, \mathcal Y)} = \underline
p^S_{(\mathcal X, \mathcal Y)} + \underline p^B_{(\mathcal X, \mathcal Y)}$, where 
\begin{equation}\label{eq:scaling}
\left|
\begin{aligned}
    &\underline p^S_{(\mathcal X, \mathcal Y)} = \frac{1}{4\pi^2}\int_{-\pi}^{\pi} dk_1 dk_2\,
    |\tilde \psi(k_1, k_2)|^2 \Theta[\epsilon'_{k_1}\in \mathcal X]\Theta[\epsilon'_{k_2}\in
    \mathcal Y]\theta(v_{k_2}-v_{k_1}), \text{ and}\\[10pt]
   & \underline p^B_{(\mathcal X, \mathcal Y)}= \frac{1}{2\pi} \int_{-\pi}^\pi d K\,
   (1-e^{-2\eta(K)}) \Theta[E'_{ K} \in \mathcal X\cap \mathcal Y]\theta(\Delta - |\cos \frac K 2
   |).
\end{aligned}\right.
\end{equation}
In this expression we use the indicator function $\Theta[P]$ which evaluates to $1$ if $P$ is true
and $0$ otherwise. It can always be expressed as a product of Heaviside step functions.

The following limits have a very simple interpretation: the probability density for the positions
of the scattering state corresponds to an interference pattern between two semi-classical
trajectories $x_k(t) = \epsilon_k' t$. Indeed, recall that $|\tilde \psi(k_1, k_2)|^2= |1+S(k_1, k_2)
e^{-i(k_1 -k_2)}|^2$. The first term corresponds to trajectories without crossing, whereas the
second one corresponds to those with an exchange (and thus a crossing) of the two particles. In
this second configuration the wavefunction acquires a phase $S(k_1, k_2)e^{-i(k_1-k_2)}$. 

In contrast, bound states become point-like in the scaling limit, following the semi-classical
trajectory $x_K(t) = E'_Kt$. Their contribution to the expectation value of the projector
$\Pi_X^Y$ stays non-zero in the scaling limit only if $X$ and $Y$ have a non-zero intersection.
\\

\paragraph*{}
The latter result can directly be used to compute the scaling limit of the one- and two-point
function of the transverse magnetisation, which read

\begin{equation}\label{eq:magnetisation_limit}
\left|
\begin{aligned}
    &\underline{\mathcal M}(\xi) =  p^B+ \int_{-\pi}^\pi \frac{dk_1 dk_2}{8\pi^2} |\tilde \psi(k_1, k_2)|^2\sgn(v_{k_1}-\xi)\sgn(v_{k_2}-\xi) \text{ and}\\
    & \underline{\mathcal{M}}(\xi_1, \xi_2) = p^B + \int_{-\pi}^{\pi}\frac{dk_1
    dk_2}{8\pi^2}|\tilde \psi(k_1, k_2)|^2
    \sgn(v_{k_1}-\xi_1)\sgn(v_{k_2}-\xi_1)\sgn(v_{k_1}-\xi_2)\sgn(v_{k_2}-\xi_2).
\end{aligned} 
\right.
\end{equation}
From these formulas it is evident that the local perturbation leads to a macroscopic deformation
of the magnetisation profile, as the latter is affected by an $\mathcal{O}(1)$ correction across
the entire light-cone of size $\sim 2t$. An interesting aspect of interactions in this setting is
that the non-locality of the elementary excitations amplifies quantum interferences to ballistic
scales. Therefore, the scattering phase becomes essentially ``readable" directly from the
magnetisation profile.
We can also see from the expression of the two-point function that it does not factorise into a
product of one-point functions, even though two different ray variables $\xi_1$ and $\xi_2$
correspond to a physical distance $(\xi_1 -\xi_2)t$ that diverges linearly in time. Even though
the initial state is a trivial product state, the perturbation thus induces a breaking of
clustering properties. The most important way the interaction affects this picture is through the
creation of bound states, which do not participate to the breaking of clustering. 

\begin{figure}[t]
    \centering
    \includegraphics[width=0.3\linewidth]{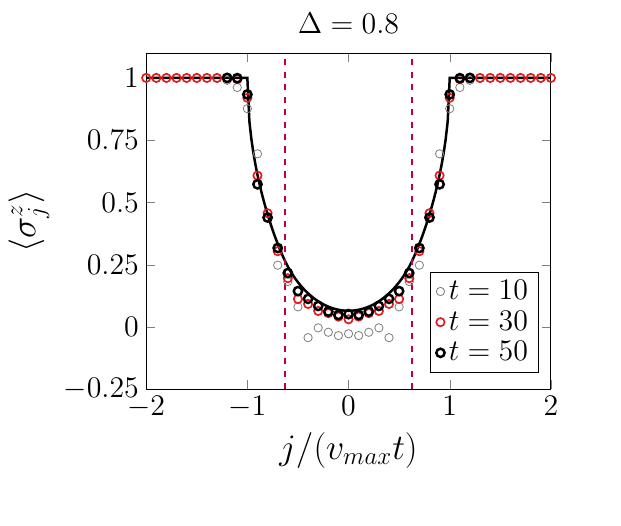}
    \includegraphics[width=0.3\linewidth]{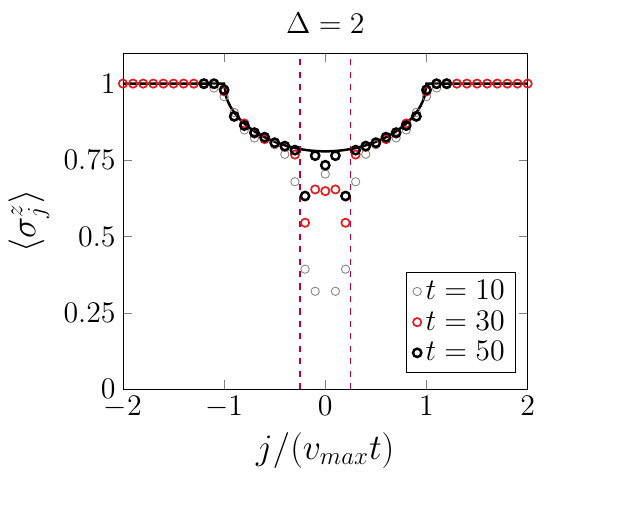}
    \includegraphics[width=0.3\linewidth]{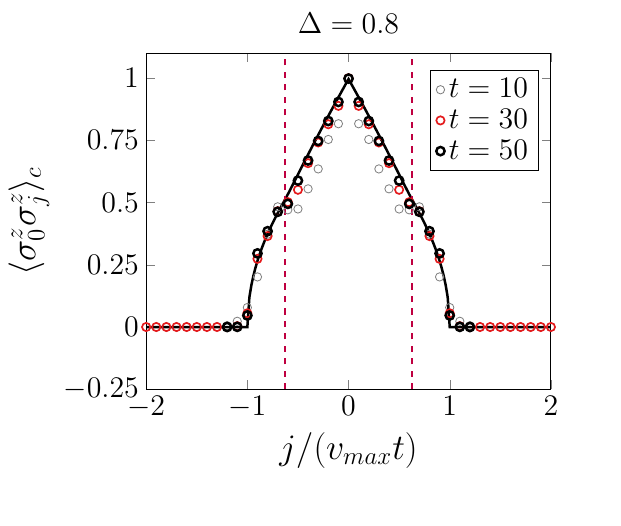}
    \caption{Magnetisation profile and connected correlations after a spin flip at site $l=0$ on
    the state $\ket{\Uparrow}$ for different values of $\Delta$. Points correspond to exact values
    of the magnetisation while the solid curve represents the asymptotic limit of
    Eq.\eqref{eq:magnetisation_limit}. The dashed line represent the
    maximal bound state velocity $v^B_{max}=1/(2\Delta)$ for $\Delta \ge 1/\sqrt 2$.}
    \label{fig:mag_prof}
\end{figure}

In FIG.~\ref{fig:mag_prof}, we compare this prediction with
exact numerical data. The agreement is already excellent
at early times, except in the region
$(-v_B^{\text{max}},v_B^{\text{max}})$ corresponding to the
light cone of the bound state, where an $\mathcal{O}(t^{-1})$
correction becomes very visible as $\Delta$ increases.
Indeed, the bound-state contribution to the magnetisation is
\begin{equation}
    \bra{\nu(t)}\sigma_j^z\ket{\nu(t)}
    =p^B-2\sum_{f_1<j<f_2}|\nu_{f_1,f_2}(t)|^2.
\end{equation}
The second term is negative and counts configurations in
which the bound state covers site $j$. Physically, the bound state
is a microscopic reversed domain, producing the observed
dip. As the bound state spreads ballistically, this correction
vanishes as $1/t$ inside its propagation
region. In the center panel, its slower spreading makes
the correction more pronounced and visible for longer times.
This disagreement is a finite-time effect, since the finite
width of the bound state becomes negligible
on ballistic scales as $t\to\infty$.

\section{Scaling limit of the Reduced Density matrix}\label{sec:rdm}
\subsection{Exact RDM}
In the rest of the paper, we aim to compute quantities that require the knowledge of the reduced
density matrix for subsystems of neighbouring spins. We consider the subsystem $A = \llbracket l,
r \rrbracket$, and also introduce the ``dual subsystem" corresponding to domain wall positions,
$\tilde A = \llbracket l -\frac 1 2, r + \frac 1 2 \rrbracket$, see
Fig~\ref{fig:protocol}. We shall from now on always indicate intervals of domain wall positions with a tilde. The partial trace of $\ket{l_1 l_2}\bra{m_1 m_2}$ is non-zero only if the
domain wall configurations in $\tilde A^c$, i.e. the complement of $\tilde A$, are identical for the bra and the ket:
 
\begin{equation}\label{eq:partial_trace_unitvec}
    \Tr_{A^c}(\ket{l_1 l_2}\bra{m_1 m_2}) = 
      \ket{l_1 l_2}_A\bra{m_1 m_2}_A t^A_{l_1 m_1}t^A_{l_2 m_2}.
\end{equation}
Here, $\ket{l_1 l_2}_A$ represents the state in the subsystem $A$ in presence of two domain walls
at positions $l_1$ and $l_2$:
\begin{equation}
    \ket{l_1 l_2}_A = \bigotimes_{j\in A}\left(\Theta[j\in (l_1,
    l_2)]\ket{\downarrow}+\Theta[j\notin (l_1, l_2)]\ket{\uparrow} \right),
\end{equation}
and the matrix $t^A_{ij}$ enforces the domain walls outside $\tilde A$ to match:
\begin{equation}
    t^A_{ij} = 
    \begin{cases} 
        1 & \text{if } i, j \in \tilde A, \\
        \delta_{ij} & \text{otherwise}.
    \end{cases}
\end{equation}

Using this formalism, we can explicitly write the RDM $\rho_A(t) =
\Tr_{A^c}(\ket{\Psi(t)}\bra{\Psi(t)})$. Dropping the explicit time dependence of the wavefunction
and density matrix,

\begin{equation}\label{eq:rdm_exact}
    \rho_A = \sum_{l_1<l_2, m_1<m_2} \Psi_{l_1 l_2}\Psi^*_{m_1 m_2} \Tr_{A^c}(\ket{l_1
    l_2}\bra{m_1 m_2}) = \sum_{l_1<l_2, m_1<m_2} \Psi_{l_1 l_2}\Psi^*_{m_1 m_2}
    \ket{l_1 l_2}_A\bra{m_1 m_2}_A t^A_{l_1 m_1}t^A_{l_2 m_2}.
\end{equation}

\subsection{Scaling limit}
In the scaling limit $|l|, |r|, t\to \infty$ with $l/t$ and $r/t$ fixed, the RDM admits a
simplified structure in the sense that both its eigenvalues and the matrix elements of products of local 
odd operators between its eigenvectors admit a scaling limit.

In order to see this, we separate the domain wall positions between three regions: $\tilde{A} =
\llbracket l - \frac{1}{2}, r + \frac{1}{2}\rrbracket$, the left complement $\tilde{L} =
\llbracket -\infty, l - \frac{3}{2}\rrbracket$, and the right complement $\tilde{R} = \llbracket r
+ \frac{3}{2}, +\infty \rrbracket$. Since $t^A_{ij}$, controlling the partial trace, is
block-diagonal between these three regions, the RDM naturally decomposes between contributions
$(X, Y)$ representing the positions of the two domain walls:

\begin{equation}\label{eq:rdm_sectors}
    \rho_A = \sum_{X, Y\in \{\tilde L, \tilde A, \tilde R\}}\rho_{A(X, Y)},
\end{equation}

where $\Tr(\rho_{A(X, Y)}) = p_{(X, Y)}$ is the probability to find a domain wall in $X$ and the other in $Y$, see
Eq.~\eqref{eq:defproba}, and the density matrix associated to this configuration is defined as:

\begin{equation}
    \rho_{A(X, Y)} = \Tr_{A^c}(\Pi_X^Y \ket{\Psi}\bra{\Psi}\Pi_{X}^Y).
\end{equation}

We will refer to $\rho_{A(\tilde L, \tilde R)}+\rho_{A(\tilde L, \tilde
L)}+\rho_{A(\tilde R, \tilde R)}$ as the $0-$particle sector, $\rho_{A(\tilde L, \tilde
A)}+\rho_{A(\tilde A, \tilde R)}$ as the $1-$particle sector and $\rho_{A(\tilde A, \tilde A)}$ as
the $2-$particle sector. Technically, these "sectors" are not orthogonal as they share the states with a domain wall at sites $l-1/2$ and $r+1/2$,
but this overlap is negligible in the scaling limit.

\subsubsection{Pure and trivial sectors}
Several sectors simplify immediately: the $0-$particle sector simply corresponds to a statistical
mixture of the two SSB vacua. If both domain walls belong to the same complement, $A$ is in the
same vacuum as the initial state before the perturbation. If instead they are on both sides of $A$
the subsystem is effectively ``flipped":

\begin{equation}
    \frac{\rho_{A(\tilde L, \tilde L)}}{p_{(\tilde L, \tilde L)}}= \frac{\rho_{A(\tilde R, \tilde R)}}{p_{(\tilde R, \tilde R)}}=\ket{\Uparrow}_A\bra{\Uparrow}_A
    \text{ and }\frac{\rho_{A(\tilde L, \tilde R)}}{p_{(\tilde L, \tilde R)}} = \ket{\Downarrow}_A \bra{\Downarrow}_A.
\end{equation}
Furthermore, the $2-$particle sector remains pure after taking the partial trace:
\begin{equation}
    \rho_{A(\tilde A, \tilde A)} = p_{(\tilde A, \tilde A)}\ket{\Psi_A}\bra{\Psi_A}, \quad \text{
    with } \ket{\Psi_A} = \frac{1}{\sqrt{p_{(\tilde A, \tilde A)}}}\sum_{l_1<l_2\in \tilde
    A}\Psi_{l_1, l_2}\ket{l_1 l_2}_A.
\end{equation}
For any internal regions $\tilde X, \tilde Y\subset \tilde A$, $\bra{\Psi_A}\Pi_{\tilde X}^{\tilde Y}\ket{\Psi_A}$ is given by the
probability ratio 
\begin{equation}\label{eq:exp_psiA}
    \bra{\Psi_A}\Pi_{\tilde X}^{\tilde Y}\ket{\Psi_A} = \frac{p_{(\tilde X, \tilde Y)}}{p_{(\tilde A, \tilde A)}},
\end{equation}
which ensures that the expectation value of any product of local odd operators in $\ket{\Psi_A}$
admits a scaling limit.

\subsubsection{Diagonalisation of the $1-$ particle sector}
\begin{figure}[t]
    \centering
    \includegraphics[width=0.3\linewidth]{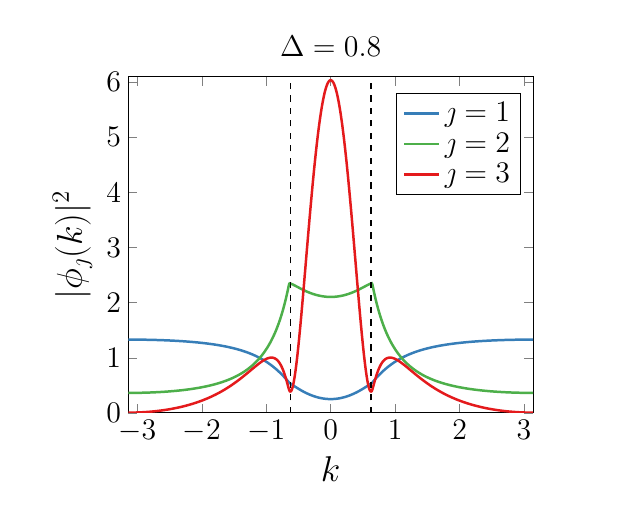}
    \includegraphics[width=0.3\linewidth]{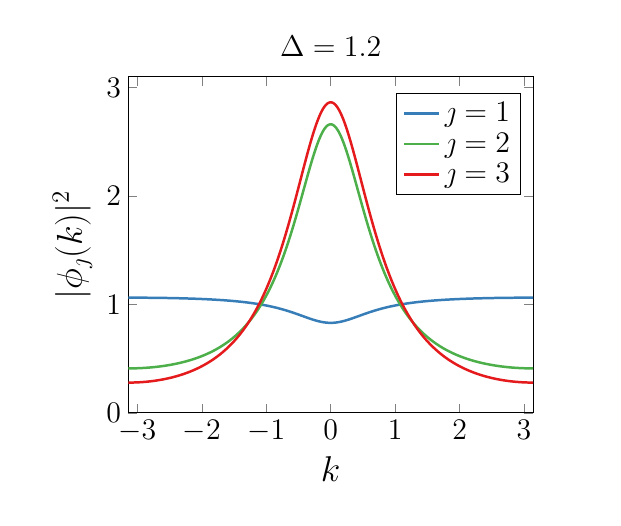}
    \includegraphics[width=0.3\linewidth]{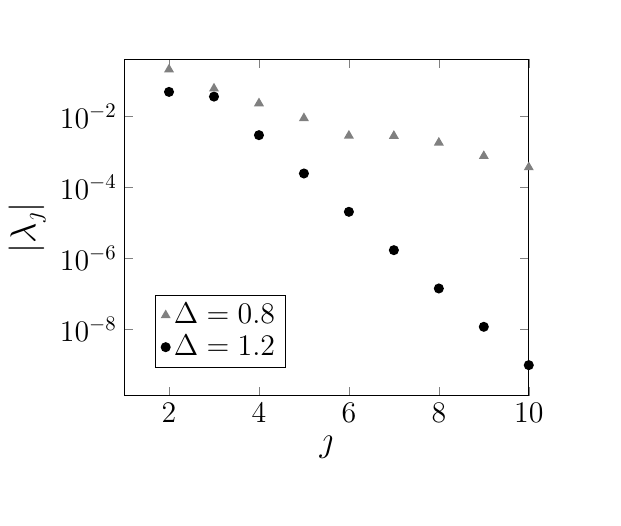}
    \caption{Numerical diagonalisation of the scattering phase using its truncated Fourier
    representation $S_{nm}$ with $|n|, |m|\le 250$. For $\Delta = 0.8$, $\sum_\jj \lambda_\jj
    \approx -0.996611$ and $\sum_\jj \lambda_\jj^2 \approx 0.999991$, while for $\Delta = 1.2$
    $\sum_\jj \lambda_\jj \approx -1$ and $\sum_\jj \lambda_\jj^2 \approx 1$ up to at least
    $10^{-8}$ precision. Left and center: representation of the first modes of the scattering
    phase. The dashed lines represent the momenta $\pm \cos^{-1}\Delta$ at which the scattering
    phase becomes non-analytic. Right: eigenvalues of the truncated Fourier representation of the
    scattering phase.}
    \label{fig:diag_S}
\end{figure}

Complications arise from the terms $\rho_{A(\tilde L, \tilde A)}$ and $\rho_{A(\tilde A, \tilde
R)}$, that have a more complex structure:
\begin{align}\label{eq:rhoA1_complicated}
    \rho_{A(\tilde L, \tilde A)} &\simeq \sum_{\substack{l_1\in \tilde L\\l_2, m_2\in \tilde
    A}}\psi_{l_1,l_2}\psi^*_{l_1 m_2}\ket{l_2}_A\bra{m_2}_A \text{ and}\\
    \rho_{A(\tilde A, \tilde R)} &\simeq \sum_{\substack{l_1, m_1\in \tilde A\\l_2\in \tilde
    R}}\psi_{l_1l_2}\psi^*_{m_1 l_2}P_A\ket{l_1}_A\bra{m_1}_A P_A,
\end{align}
where we define $\ket{m}_A = \ket{l-\frac 1 2, m}_A$, $P_A \ket{m}_A = \ket{m, r+\frac 1 2}_A$. In
this expression $\psi_{l_1,l_2}$ is the scattering states wavefunction, see
Eq.~\eqref{eq:magnons}. We dropped the bound states contribution that vanishes in the scaling
limit (see Eq.~\eqref{eq:scaling}) since $\tilde L$, $\tilde A$ and $\tilde R$ have no overlap.
The effective number of terms in the sum over $\tilde L$ (resp. $\tilde R$) diverges linearly in
time as the light-cone expands. On the other hand, each individual term vanishes like $1/t$,
making the scaling non-obvious. To resolve this issue, we recall the integral representation of
$\psi_{l_1 l_2}$:
\begin{equation}
\psi_{f_1, f_2} = \frac{1}{4\pi^2}\int_{-\pi}^\pi dk_1 dk_2\, \tilde \psi(k_1, k_2) e^{i[k_1
(f_1-1/2) + k_2 (f_2+1/2)]}e^{-i(\epsilon_{k_1}+\epsilon_{k_2})t}.
\end{equation}
The coupling between the two domain walls is entirely contained within the function $\tilde
\psi(k_1, k_2) =e^{i(k_1-k_2)}+S(k_1, k_2)$. Since the scattering phase satisfies $S(k_1, k_2) =
S(k_2, k_1)^*$, it can be seen as the Kernel of a Hermitian operator acting on the compact support
$(-\pi, \pi)$. By the Hilbert-Schmidt theorem, it can be decomposed as 

\begin{equation}\label{eq:S_diag}
    S(k_1, k_2) = \sum_{\jj=1}^{+\infty} \lambda_\jj \phi_\jj(k_1)\phi_\jj^*(k_2),
\end{equation}
where the set of eigenvalues $\lambda_\jj$ is discrete, accumulating only around $0$ and the
eigenfunctions $\phi_j$ are orthonormal: $\int_{-\pi}^\pi dk\, \phi_{\jj}(k)\phi^*_{\jj'}(k) =
2\pi \delta_{\jj \jj'}$. 
The left and middle panels of FIG~\ref{fig:diag_S} illustrate the momentum dependence of the first three eigenfunctions $\phi_\jj$ for $\Delta=0.8$ and $\Delta=1.2$. 
For $\Delta<1$, these profiles display a non-analyticity near the momenta $k=\pm\arccos\Delta$, whereas they are smooth for $\Delta>1$.

The dominant terms of Eq.~\eqref{eq:S_diag} can be very well approximated numerically by
diagonalising the truncated matrix of Fourier coefficients of $S(k_1, k_2)$, given in appendix
\ref{ap:S_diag}. The properties of the scattering phase,
\begin{equation}
    S(k, k) = -1 \text{ and } S(k_1, k_2) S(k_2, k_1) = 1,
\end{equation}
impose the eigenvalues to satisfy
\begin{equation}
    \sum_{\jj=1}^{+\infty} \lambda_\jj = -1 \text{ and }\sum_{\jj=1}^{+\infty} \lambda_\jj^2 = 1.
\end{equation}

These identities provide a robust consistency check of the numerical diagonalisation. To recover
the full expression of $\tilde \psi(k_1, k_2)$, we define $\lambda_0 = 1$ and $\phi_0(k) = e^{i
k}$ in such a way that $\tilde \psi(k_1, k_2) = \sum_{\jj = 0}^{+\infty} \lambda_\jj \phi_\jj(k_1)
\phi_\jj^*(k_2)$. Furthermore, we introduce the $\tilde X$-restricted Gram matrices of the modes
$\phi_\jj$ as:
\begin{align}
    &G_{\jj \jj'}(\tilde X) = \sum_{l\in \tilde X}\left( \frac{1}{4\pi^2}\int_{-\pi}^{\pi}dk\, dp
    \, \phi_\jj(k) \phi^*_{\jj'}(p) e^{i(k-p)(l-1/2)} e^{-i(\epsilon_{k}-\epsilon_{p})t}\right)\\
    &G^{c}_{\jj \jj'}(\tilde X) = \sum_{l\in \tilde X}\left( \frac{1}{4\pi^2}\int_{-\pi}^{\pi}dk\,
    dp \, \phi^*_{\jj'}(k) \phi_{\jj}(p) e^{i(k-p)(l+1/2)}
    e^{-i(\epsilon_{k}-\epsilon_{p})t}\right),
\end{align}
where $\tilde X$ is a subset of domain wall positions.

These definitions allow to recast Eq.~\eqref{eq:rhoA1_complicated} as a discrete sum over the
eigenvalues $\lambda_\jj$, which typically decay exponentially, see Fig~\ref{fig:diag_S}. If we
focus on $\rho_{A(\tilde L, \tilde A)}$, we get
\begin{equation}
    \rho_{A(\tilde L, \tilde A)}= \sum_{\jj, \jj'=0}^{+\infty} \lambda_\jj \lambda_{\jj'} G_{\jj
    \jj'}(\tilde L)\ket{\phi_\jj}_A\bra{\phi_\jj'}_A.
\end{equation}
The states $\ket{\phi_j}_A$ are the lattice representation of the eigenfunction $\phi_j(k)$ on the
subsystem $A$:
\begin{equation}
    \ket{\phi_\jj}_A = \sum_{l\in \tilde A} \frac{1}{2\pi}\int_{-\pi}^\pi dk \phi_j^*(k) e^{i
    k(l+1/2)}e^{-i\epsilon_k t}\ket{l}_A, \quad \braket{\phi_\jj}{\phi_{\jj'}}_A = G^{c}_{\jj
    \jj'}(\tilde A).
\end{equation}

This framework reduces the problem of finding the eigenvectors and eigenvalues of $\rho_A$ to that
of diagonalising the matrix $M^{(\tilde L, \tilde A)} = \Lambda G(\tilde L) \Lambda (G^c(\tilde
A))$, where $\Lambda = \diag (\{\lambda_j\})$. In the scaling limit, the Gram matrices $G(\tilde
X)$ and $G^c(\tilde X)$ converge to the overlap of the eigenfunctions $\phi_\jj(k)$ restricted to
the ray-interval $\mathcal{X}$:
\begin{equation}
\underline G_{\jj \jj'}(\mathcal X) = (\underline G_{\jj' \jj}^c(\mathcal X))^* = \frac{1}{2\pi}
\int_{-\pi}^\pi dk \, \phi_\jj(k) \phi_{\jj'}^*(k) \Theta[\epsilon'_k \in \mathcal{X}].
\end{equation}
The existence of this limit ensures that the spectrum of $\rho_{A(\tilde L, \tilde A)}$ admits a
well-defined scaling limit, namely:
\begin{equation}\label{eq:rdm_eigvals_scaling}
    \{\underline \mu_j^{\mathcal A(\mathcal L, \mathcal A)}\} = \Spec [\Lambda \underline
    G(\mathcal L) \Lambda \underline G^\dagger (\mathcal A)].
\end{equation}
Furthermore, the eigenvectors of $\rho_{A(\tilde L, \tilde A)}$ are given by a linear combination
of the states $\ket{\phi_{\jj'}}_A$, 
\begin{equation}\label{eq:rdm_eigenstates}
    \ket{\varphi_n^{A(\tilde L, \tilde A)}} = \sum_{j} W_{nj}^{(\tilde L, \tilde A)} \ket{\phi_j}_A,
\end{equation}
where the coefficients $W_{nj}^{(\tilde L, \tilde A)}$ are the eigenvectors of $M^{(\tilde L,
\tilde A)}$. These coefficients converge to stable values $\underline W_{nj}^{(\mathcal L,
\mathcal A)}$ in the scaling limit, namely the eigenvectors of $\underline M^{(\mathcal L,
\mathcal A)} = \Lambda \underline G(\mathcal L) \Lambda \underline G^\dagger (\mathcal A)$. 

\section{Entanglement asymmetry}\label{sec:ent_asym}
Having access to the reduced density matrix $\rho_A$, we can now proceed to study the quantum
coherence associated with the fluctuations of the transverse magnetisation $Z_A = \sum_{\jj\in
A}\sigma^z_\jj$ within the subsystem. To do so, we first focus on the entanglement asymmetry of
$\rho_A$ with respect to $Z_A$:
\begin{equation}
    \Delta S(\rho_A, Z_A) = S(\mathcal G_{Z_A}(\rho_A))-S(\rho_A).
\end{equation}
This quantity is defined as the relative entropy between the state $\rho_A$ and its ``twirled''
version,
\begin{equation}
    \mathcal G_{Z_A}(\rho_A) = \sum_{z\in \Spec Z_A} \Pi_{z}\,\rho_A\, \Pi_{z}.
\end{equation}
It is expressed in terms of the projector $\Pi_z$ on the eigenspace of $Z_A$ with eigenvalue $z$.
The twirled state $\mathcal G_{Z_A}(\rho_A)$ represents the state of $A$ after a non-selective
measurement of $Z_A$: physically, this operation removes all off-diagonal elements - corresponding
to coherence - between charge sectors. Consequently, while the fluctuations of $Z_A$ in $\rho_A$
and $\mathcal G_{Z_A}(\rho_A)$ are identical, those in the twirled state can be interpreted as
solely due to classical uncertainty. 

$\Delta S(\rho_A, Z_A)$ is an asymmetry measure with respect to the $U(1)$ group generated by
$Z_A$. While the latter is not a symmetry of the Hamiltonian, $[H, Z]\neq 0$, it is a symmetry of the SSB ground
state, which has fixed magnetisation. The Entanglement Asymmetry (EA) satisfies two important
properties:
(i) it is positive \cite{Nielsen2010Computation}, $\Delta S(\rho_A, Z_A)\ge 0$, and
(ii) it vanishes precisely when the state is symmetric with respect to the charge
\cite{Ares2023Entanglement}, i.e., when there is no coherence between charge sectors,
\[
\Delta S(\rho_A, Z_A)=0 \quad \Leftrightarrow \quad [\rho_A, Z_A]=0.
\] 

After a local quench on the SSB ground state, one can expect the symmetry to be violated at
intermediate times as excitations propagate, and then restored at late times as the system locally
relaxes into the zero-particle sector.

Computing the exact value of $\Delta S_{Z_A}$ is generally a formidable task. However, in our
setting, we can take advantage of the low dimensionality of the $2$-particle sector to numerically
compute its exact value in a time that scales polynomially with respect to the subsystem size. 

To gain physical insight into the behaviour of $\Delta S(\rho_A, Z_A)$, we note that its leading
behaviour is relatively insensitive to the structure of the RDM. Actually, we can work out two
simple bounds to approximate it. Let $B$ be the complement of region $A$, and let us define the
twirled state of full system $\ket \Psi$ with respect to $Z_A$:
\begin{equation}
    \sigma^{(A)} = \mathcal{G}_{Z_A}(\ket \Psi \bra \Psi).
\end{equation}
Since the twirling operator $\mathcal{G}_{Z_A}$ acts non-trivially only on the subsystem $A$, it
satisfies
\begin{align}
    \Tr_B(\sigma^{(A)}) &= \mathcal{G}_{Z_A}(\Tr_B(\ket{\Psi}\bra{\Psi})) =
    \mathcal{G}_{Z_A}(\rho_A), \\
    \Tr_A(\sigma^{(A)}) &= \Tr_A\left(\sum_{z\in \Spec Z_A} \Pi_z \ket{\Psi}\bra{\Psi}
    \Pi_z\right) = \Tr_A(\ket{\Psi}\bra{\Psi}) = \rho_B.
\end{align}
The triangle inequality for entanglement entropy $S_{AB} \ge |S_A-S_B|$ applied to $\sigma^{(A)}$
yields
\begin{equation}
    S(\sigma^{(A)})\ge |S(\mathcal G_{Z_A}(\rho_A))-S(\rho_B)|.
\end{equation}
Recalling that $\ket \Psi \bra \Psi$ is pure, $S(\rho_A) = S(\rho_B)$. Therefore, the inequality
becomes:
\begin{equation}
    S(\sigma^{(A)})\ge  \Delta S(\rho_A, Z_A).
\end{equation}
Note that the entropy of $\sigma^{(A)}$ is precisely the asymmetry of the full state with respect
to $Z_A$, since $S(\ket{\Psi}\bra{\Psi})=0$.
On the other hand, the subadditivity of the Von-Neumann entropy, $S(\rho_{AB}) \le S(\rho_A) +
S(\rho_B)$, applied to $\sigma^{(A)}$ gives a lower bound:
\begin{equation}
    S(\sigma^{(A)}) \le S(\mathcal{G}_{Z_A}(\rho_A)) + S(\rho_B).
\end{equation}
Rearranging and substituting $S(\rho_A) = S(\rho_B)$, we finally obtain that the subsystem
asymmetry is bounded by the entropy of the charge fluctuations:

\begin{figure}
    \centering
    \includegraphics[width=0.4\linewidth]{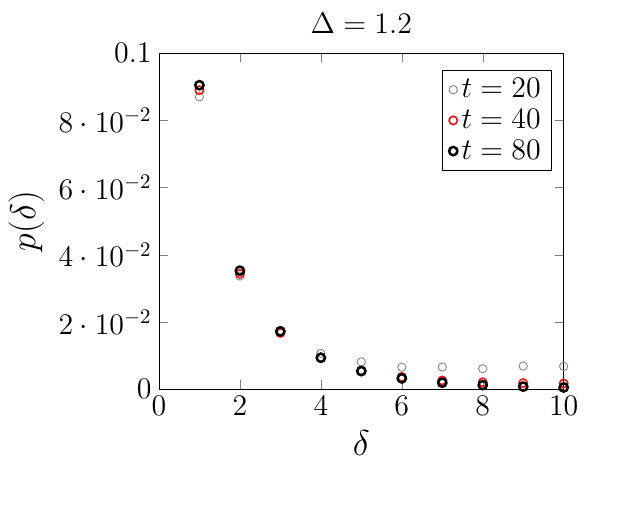}
    \includegraphics[width=0.4\linewidth]{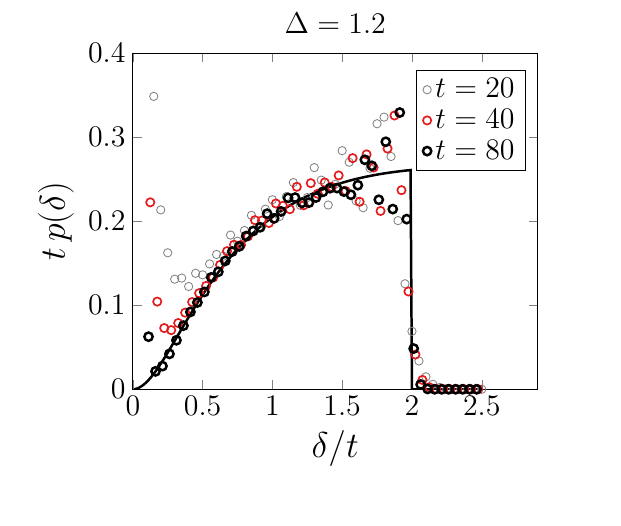}
    \caption{Probability $p(\delta)$ to find two domain walls separated by a distance $\delta$.
    Left: time-independent contribution to the probability distribution (bound states). Right:
    scaling limit of the time-dependent part of the distribution (scattering states). The
    numerical data is compared with the (coarse grained) underlying continuous distribution.}
    \label{fig:delta_distrib}
\end{figure}

\begin{equation}
    S(\sigma^{(A)})-2S(\rho_A) \le \Delta S(\rho_A, Z_A)\le S(\sigma^{(A)}).
\end{equation}
Our analysis in section \ref{sec:rdm} implies that the entanglement entropy $S(\rho_A)$ admits a
well-defined scaling limit, given by the Shannon entropy of the RDM eigenvalues $\underline
\mu_j^{\mathcal A}$:
\begin{equation}
S(\rho_A) \to -\sum_j \underline \mu_j^{\mathcal A} \log \underline \mu_j^{\mathcal A}.
\end{equation}
The eigenvalues $\underline \mu_j^{\mathcal A}$ decay rapidly (a consequence of the decay of the
Hilbert-Schmidt coefficients $\lambda_j$), so this limit is finite. Consequently, the entanglement
entropy $S(\rho_A)$ contributes only an $\mathcal{O}(1)$ term to the total asymmetry. In the
scaling limit, we can therefore approximate the entanglement asymmetry using the asymmetry
$S(\sigma^{(A)})$ of the full state $\ket{\Psi}$ with respect to the subsystem charge $Z_A$:
\begin{equation}
    \Delta S (\rho_A, Z_A) = S(\sigma^{(A)}) + \mathcal{O}(1),
\end{equation}
where the $\mathcal{O}(1)$ correction is negative or zero (between $-2S(\rho_A)$ and $0$).
This approximation is particularly useful because $S(\sigma^{(A)})$ only depends on the
probability distribution $p_{Z_A}(z)$ for the different possible magnetisations in $A$:
\begin{equation}
    S(\sigma^{(A)}) = -\sum_{z \in \Spec(Z_A)}p_{Z_A}(z)\log p_{Z_A}(z),
\end{equation}
which is much easier to compute than the exact entanglement asymmetry of $\rho_A$. The probability
of finding a magnetisation $Z_A=|A|-2d$ (corresponding to $d$ flipped spins) is given by 
\begin{multline}
    p_{Z_A}(|A|-2d) = \delta_{d, 0}(p_{(\tilde L, \tilde L)}+p_{(\tilde R, \tilde R)}) +
    \delta_{d, |A|}p_{(\tilde L, \tilde R)}
\\+\sum_{\jmath \le l-\frac 3 2}|\Psi_{\jmath, l-\frac 1 2 + d}|^2+ \sum_{\jmath'\ge r+\frac 3
2}|\Psi_{r+\frac 1 2 -d, \jmath'}|^2+\sum_{\jmath = l-\frac 1 2}^{r+\frac 1 2 -d}|\Psi_{\jmath,
\jmath+d}|^2.
\end{multline}
The extreme values of the magnetisation, $Z_A = \pm |A|$,  carry an $\mathcal O(1)$ weight
corresponding to the configurations where both domain walls are outside the subsystem. Similarly,
the bound states give $\mathcal{O}(1)$ contributions to the near-to-maximal magnetisations, $Z_A =
|A|-2d$, that decay exponentially with $d$. 

In contrast, the scattering states give $\mathcal{O}(1/t)$ contributions spread on a support of
size $\sim 2t$ (the size of the light-cone) when at least one domain wall is inside the subsystem.
This distinction has a strong impact on the behaviour of the asymmetry for different Rényi indexes
$\alpha$. For $\alpha >1$, the sum is dominated by the $\mathcal O(1)$ weights (vacuum and bound
states), while the Von-Neumann ($\alpha=1)$ asymmetry is dominated by the $\mathcal{O}(1/t)$
contributions. Consequently, the standard Replica trick \cite{Holzhey1994Geometric,
Calabrese2004Entanglement} methods to compute the limits as $n\to 1$ of $\Tr(\rho_A^n)$ and
$\Tr(\mathcal G_{Z_A}(\rho_A)^n)$ are not adapted to this problem. Keeping only the dominant
$\mathcal{O}(1/t)$ terms, we find that the asymmetry should behave as
\begin{equation}
    \Delta S(\rho_A, Z_A) = (p^S - p^S_0(t)) \log(2t)+\mathcal{O}(1),
\end{equation}
where $p^S_0(t) = p^S_{(\tilde L, \tilde L)}(t)+p^S_{(\tilde R, \tilde R)}(t)+p^S_{(\tilde L,
\tilde R)}(t)$ is the time-dependent probability to have a scattering state where both domain
walls are outside the subsystem. This formula implies that the quantity $\Delta S(\rho_A,
Z_A)/\log (2t)$ is well defined in the scaling limit:

\begin{equation}\label{eq:lim_EA}
    \lim_{\substack{|A|, t\to \infty\\ 2t/|A| = \tau}}\frac{\Delta S(\rho_A, Z_A)}{\log (2t)} = p^S
    - \underline{p^S_0}(\tau).
\end{equation}

The logarithmic growth is a consequence of the ballistic spreading
of the domain walls. Their separation determines the
number of flipped spins, so scattering states coherently span a
number of magnetisation sectors that diverges linearly in time.
Instead, a bound state keeps a microscopic separation: translating
it within $A$ does not change the magnetisation, and its size
fluctuations only contribute an $\mathcal O(1)$ term to the asymmetry.

At late
times $t\gg |A|$, the probability $p^S_0(t)$ approaches $p^S$ as $1/t$ as the slowest
quasiparticles escape the subsystem. Consequently, the asymmetry vanishes to its vacuum value as
the symmetry is locally restored. The leading behaviour of the asymmetry is quite universal, and
is affected by interactions only through the term $p^S-\underline p^S_0(t)$ that discards the
bound states.

\begin{figure}
    \centering
    \includegraphics[width=0.4\linewidth]{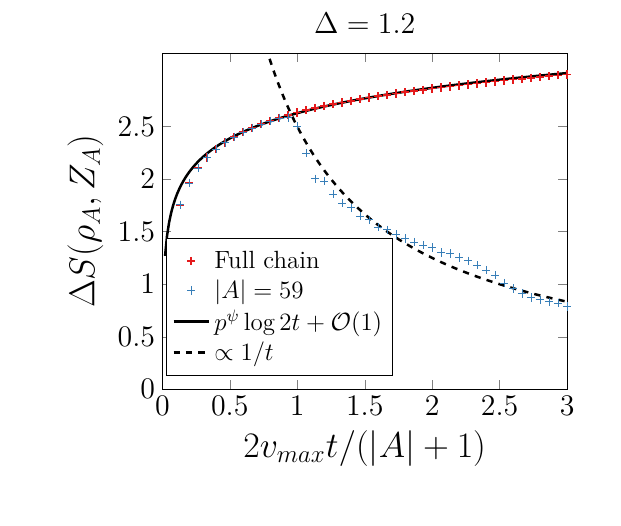}
    \includegraphics[width=0.4\linewidth]{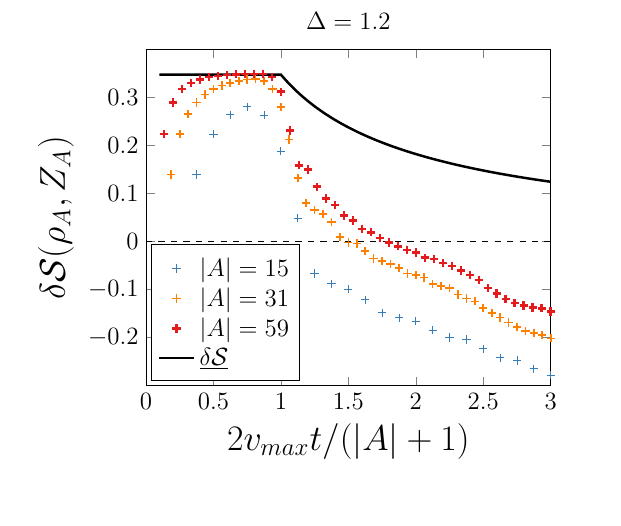}
    \caption{Left: Time evolution of asymmetry after a spin flip. Right: Comparison between the
    asymptotic prediction and the numerical values of $\delta \mathcal S(\rho_A, Z_A) = (\Delta
    S(\rho_A, Z_A)-\Delta S_0)/\log t$ for different subsystem sizes.}
    \label{fig:asym}
\end{figure}

While Eq.~\eqref{eq:lim_EA} captures the leading-term behaviour of the asymmetry, in practice the
$\mathcal O(1)$ correction is comparable with the logarithmic one for the times and subsystem
sizes we study numerically. At intermediate times $t\le|A|/2$, the subsystem is effectively pure
as the quasiparticles from the initial perturbation have not yet reached the subsystem boundaries.
In this regime the $\mathcal O(1)$ correction can be very well approximated by a constant term
$\Delta S_0$, taking into account the contribution of the bound state (see left panel of
Fig.~\ref{fig:delta_distrib}) and the differential entropy associated with the non-oscillating
part of the probability distribution for the magnetisation (right panel of
Fig.~\ref{fig:delta_distrib}). Taking into account this correction allows for a very good fit of
the numerical data, see Fig~\ref{fig:asym}. Conversely, both the bound state and scattering states
contributions to the non-vacuum part of the RDM can be expanded at late times as corrections to
the infinite time (vacuum) limit. The probability to find a domain wall or a bound state in $A$
decays as $1/t$, which preserves the predicted $1/t$ power-law decay of the asymmetry.

\section{Quantum Fisher Information}\label{sec:qfi}
The entanglement asymmetry, discussed in the previous section, provides a rigorous measure of the
coherence between the different magnetisation sectors. However, it effectively ``counts" this
number of sectors independently of the magnetisation associated to each of them. To further
characterise the quantum fluctuations associated to coherence between different magnetisation
sectors, it is natural to extend this study to the computation of the Quantum Fisher Information
(QFI), which is sensitive to the magnitude of the fluctuations between the magnetisation sectors.
We will actually work with the normalised Quantum Fisher Information (nQFI)  $\chi(\rho, O) = F_Q(\rho, O)/(4||O||^2)$, as used in Ref~\cite{ferro2025kickingquantumfisherinformation}. 
Here $F_Q(\rho, O)$ is the QFI for the unitary transformation $\rho_\theta = e^{-i\theta O} \rho e^{i\theta O}$
, which is defined for a density matrix $\rho =
\sum_i \mu_i \ket i \bra i$ as 

\begin{equation}\label{eq:qfi_def}
    F_Q(\rho, O) = 2\sum_{i, j=1}^{+\infty}\frac{(\mu_i-\mu_j)^2}{\mu_i +
    \mu_j}|\bra{i}O \ket{j}|^2. 
\end{equation}

The reason we are specifically interested in the nQFI is its ability to detect the physical scale
of quantum correlations, and in particular the breaking of clustering due to quantum
superposition. To formalise this, we rely on the following properties of the nQFI:
\paragraph{Link with variance:} For a pure state and this unitary transformation the QFI reduces to $4$ times the variance:
\begin{equation}
F_Q(\ket \psi \bra \psi,O)=4\operatorname{Var}(\ket \psi \bra \psi, O),
\qquad\chi(\ket \psi \bra \psi, O) = \frac{\Var(\ket \psi \bra \psi, O)}{||O||^2}.
\end{equation}
\paragraph{Sensitivity to pure coherence:} unlike the variance, the (n)QFI is convex:
\begin{equation}
    \chi(\sum_i p_i \rho_i)\le \sum_i p_i \chi(\rho_i),
\end{equation}
and can actually be identified (up to a constant factor) with the convex roof of the variance
\cite{yu2013quantum,Toth2013Extremal}. This property ensures that, while the variance is sensitive to both
classical and quantum fluctuations of $O$, the (n)QFI is only sensitive to quantum ones.
\paragraph{Normalisation:} nQFI satisfies the bounds
\begin{equation}
    0\le \chi(\rho, O)\le 1.
\end{equation}
$\chi(\rho, O) = 0$ for a symmetric state ($[\rho, O] = 0$). For an extensive observable $O=\sum_j o_j$ made of local operators,
such as $Z_A$, this maximal value is attained by a cat-like state, i.e. an equal-weight superposition of two states with minimal
($-||O||$) and maximal ($||O||$) values of $O$.

For these reasons, nQFI is able to detect a breakdown of clustering properties due to quantum
superposition. In states with clustering, connected correlation functions $\langle o_i
o_j\rangle_c = \langle o_i o_j\rangle-\langle o_i\rangle \langle o_j\rangle$ tend to $0$ as the
distance $|i-j|$ increases. In this setting, even in the presence of algebraically decaying
correlations the variance of an extensive operator $\Var(\rho_A, O_A)$ scales like some power of
its volume, $|A|^\alpha$, with $\alpha <2$. This implies
\begin{equation}
    \chi (\rho_A, O_A)\le \frac{\Var(\rho_A, O_A)}{||O_A||^2} \underset{|A|\to \infty}\to 0.
\end{equation}

Consequently, a non-zero value of $\chi$ in the scaling limit signals a violation of clustering at
ballistic scales: the state must be a coherent superposition of magnetisation sectors whose total
charges differ by an amount proportional to the system size. This effect was notably observed
after a local quench on the SSB ground state of the Ising model in the ferromagnetic phase
\cite{ferro2025kickingquantumfisherinformation}, or after joining a SSB ground state with a
thermal one in the growing interface between order and disorder \cite{maric2025disorder,
maric2025macroscopic}. This section aims to generalise the results of
Ref.~\cite{ferro2025kickingquantumfisherinformation} to the interacting case and to introduce a
Bethe-Ansatz method for computing the nQFI after local quenches. 

Let $\underline{\chi}_{\mathcal{A}}$ be the value of the normalised QFI in the limit $|A|, t \to
\infty$ with the ratio $\tau = 2t/|A|$ held constant (we will see that such a limit exists).
Taking advantage of decomposition~\eqref{eq:rdm_sectors} of the RDM: $\rho_A = \sum_{(X, Y)}\rho_{A(X, Y)}$, and noting that the magnetisation operator $Z_A$ does not mix the different
sectors $(\tilde X, \tilde Y)$ in the scaling limit, the nQFI can be decomposed into the sum of
the contributions from each sector of the RDM:
\begin{equation}
    \underline \chi_{\mathcal A} = \sum_{(\mathcal X, \mathcal Y)}\underline \chi_{\mathcal A(\mathcal X, \mathcal Y)}.
\end{equation}
Here, $\underline{\chi}_{\mathcal{A}(\mathcal X, \mathcal Y)}$ represents the scaling limit of the
nQFI for the density matrix $\rho_{A(X, Y)}$. Since the sectors $(\mathcal{L}, \mathcal{L})$,
$(\mathcal{R}, \mathcal{R})$, and $(\mathcal{L}, \mathcal{R})$ correspond to vacuum states, their
contributions to the QFI are identically $0$ (the variance of $Z_A$ vanishes). The $(\mathcal A,
\mathcal A)$ sector - where both domain walls are inside the subsystem - is pure, therefore its
contribution to the QFI is simply given by a variance. The calculation is done easily by
integrating the scaling limit of $\bra{\Psi_A}\sigma^z_i \sigma^z_j\ket{\Psi_A}_c$, given by
Eq.~\eqref{eq:exp_psiA}. The result takes the simple form: 

\begin{multline}
    \underline \chi_{\mathcal A (\mathcal A, \mathcal A)} = \tau^2\left[\int_{-\pi}^\pi \frac{dk_1
    dk_2}{8\pi^2} |\tilde \psi(k_1, k_2)|^2
    (\epsilon'_{k_1}-\epsilon'_{k_2})^2\Theta[\epsilon'_{k_1}\in \mathcal
    A]\Theta[\epsilon'_{k_2}\in \mathcal A]\right.\\
    \left.-\left(\int_{-\pi}^\pi \frac{dk_1 dk_2}{8\pi^2} |\tilde \psi(k_1, k_2)|^2
    |\epsilon'_{k_1}-\epsilon'_{k_2}| \Theta[\epsilon'_{k_1}\in \mathcal
    A]\Theta[\epsilon'_{k_2}\in \mathcal A]\right)^2 \bigg /\underline p_{(\mathcal A, \mathcal A)}\right].
\end{multline}

The sectors $(\mathcal L, \mathcal A)$ and $(\mathcal A, \mathcal R)$, where one domain wall is
traced out, present the main difficulty as they are mixed states. 

The purity of $\rho_{A(\tilde L,\tilde A)}$ (which is by reflection symmetry equal to that of $\rho_{A(\tilde A,\tilde R)}$ when $A$ is symmetric around the origin) is obtained from
a stationary phase approximation as
\begin{equation}
\begin{split}
\operatorname{Tr}\!\left[\rho_{A(\tilde L,\tilde A)}^2\right]
\longrightarrow{}&
\int_{-\pi}^{\pi}\frac{dk_2\,dp_2}{(2\pi)^2}\,
\Theta[\epsilon'_{k_2}\in\mathcal A]\,
\Theta[\epsilon'_{p_2}\in\mathcal A]
\left|
\int_{-\pi}^{\pi}\frac{dk_1}{2\pi}\,
\tilde\psi(k_1,k_2)\tilde\psi^*(k_1,p_2)\,
\Theta[\epsilon'_{k_1}\in\mathcal L]
\right|^2 .
\end{split}
\label{eq:one_particle_purity}
\end{equation}
Numerical evaluation of this integral, see the middle panel of FIG.~\ref{fig:qfi}, shows that the corresponding state is nearly pure for sufficiently large values of $\Delta$, at least at intermediate times when the probability to be in the one particle sector is most significant.

Therefore, the nQFI is well approximated by the
variance of $Z_A$ in these states, with a rigorous bound given by the purity: for a normalised state $\rho$ with $\Tr \rho^2 \ge \frac 1 2$,
\begin{equation}\label{eq:qfi_pure}
     \frac{\Var (\rho, \mathcal O)}{||O||^2} - 4\left(1-\sqrt{2\Tr \rho^2-1}\right)\le \chi (\rho,
     \mathcal O)\le \frac{\Var (\rho, \mathcal O)}{ ||\mathcal O||^2 },
\end{equation}
that we derive in appendix \ref{ap:bounds}. Under this ``pure state approximation", the $(\mathcal
L, \mathcal A)$ contribution to the nQFI simplifies to 

\begin{multline}
    \underline \chi_{\mathcal A (\mathcal L, \mathcal A)}^{(\text{pure})} =
    \tau^2 \left[\int_{-\pi}^\pi \frac{dk_1 dk_2}{4\pi^2} |\tilde \psi(k_1, k_2)|^2
    (\epsilon'_{k_2})^2\Theta[\epsilon'_{k_1}\in \mathcal L]\Theta[\epsilon'_{k_2}\in \mathcal
    A]\right.\\
    \left.-\left(\int_{-\pi}^\pi \frac{dk_1 dk_2}{4\pi^2} |\tilde \psi(k_1, k_2)|^2
    \epsilon'_{k_2}\Theta[\epsilon'_{k_1}\in \mathcal L]\Theta[\epsilon'_{k_2}\in \mathcal
    A]\right)^2\bigg /\underline p_{(\mathcal L, \mathcal A)}\right].
\end{multline}
\subsubsection*{Higher-order corrections to the nQFI}

To go beyond this approximation, we assume that the eigenvalues $\mu_j^{ A(\tilde L, \tilde A)}$
of $\rho_{A(\tilde L, \tilde A)}$ are indexed in such a way that they decrease with $j$. We then
introduce the truncated reduced density matrix
\begin{equation}
    \rho_{A(\tilde L, \tilde A)}^{(n)} = \sum_{\jj= 1}^n \mu_j^{ A(\tilde L, \tilde A)}
    \ket{\varphi_\jj^{A(\tilde L, \tilde A)}} \bra{\varphi_\jj^{A(\tilde L, \tilde A)}},
\end{equation}
that only keeps the $n$ dominant eigenvectors of the RDM. The advantage of using the truncated
RDM is that using the techniques introduced in Section \ref{sec:rdm}, its nQFI $\underline
\chi_{\mathcal A(\tilde L, \tilde A)}^{(n)}$ in the scaling limit can be expressed as a finite sum
involving the matrix elements $\underline{\mathcal{Z}}^{[1]}$ and $\underline{\mathcal{Z}}^{[2]}$
(the first and second moments of the domain wall rays) over the $n$ largest eigenvalues of the RDM:

\begin{equation}\label{eq:qfi_cut}
    \underline \chi_{\mathcal A(\mathcal L, \mathcal A)}^{(n)} = \sum_{j=1}^n \underline \mu_j^{
    A(\mathcal L, \mathcal A)}  |\underline{\mathcal{Z}}^{[2]\mathcal A(\mathcal L, \mathcal
    A)}_{jj}|-2 \sum_{i, j=1}^n\frac{\underline \mu_i^{ A(\mathcal L, \mathcal A)}   \underline
    \mu_j^{ A(\mathcal L, \mathcal A)}  }{\underline \mu_i^{ A(\mathcal L, \mathcal A)}   +
    \underline \mu_j^{ A(\mathcal L, \mathcal A)}  }|\underline{\mathcal{Z}}^{[1]\mathcal
    A(\mathcal L, \mathcal A)}_{ij}|^2,
\end{equation}
with (see Eq.\eqref{eq:rdm_eigenstates})
\begin{equation}\label{eq:moments}
    \underline{\mathcal{Z}}^{[\alpha]\mathcal A(\mathcal L, \mathcal A)}_{mn} = \sum_{i, j}
    (\underline W_{ni}^{(\mathcal L, \mathcal A)})^* \underline W_{mj}^{(\mathcal L, \mathcal A)}
    \int_{-\pi}^\pi \frac{dk}{2\pi} \, \phi_i(k) \phi_{j}^{*}(k) (\tau\epsilon'_k)^\alpha
    \Theta[\epsilon'_k \in \mathcal{A}].
\end{equation}

In appendix \ref{ap:bounds}, we also derive the following bounds for the nQFI of the truncated RDM,
which ensure that the exact QFI can be approached to any level of precision simply by increasing
$n$:
\begin{equation}\label{eq:qfi_bounds_cut}
    \underline \chi_{\mathcal A(\tilde L, \tilde A)}^{(n)} -4n\underline \epsilon_n  \le
    \underline \chi_{A(\tilde L, \tilde A)} \le \underline \chi_{\mathcal A(\tilde L, \tilde
    A)}^{(n)} + \underline \epsilon_n,
\end{equation}
where $\underline \epsilon_n = \sum_{j = n+1}^{+\infty} \underline \mu_j^{ A(\mathcal L, \mathcal
A)}$ is the missing weight in the truncated RDM.

\begin{figure}
    \centering
    \includegraphics[width=0.32\linewidth]{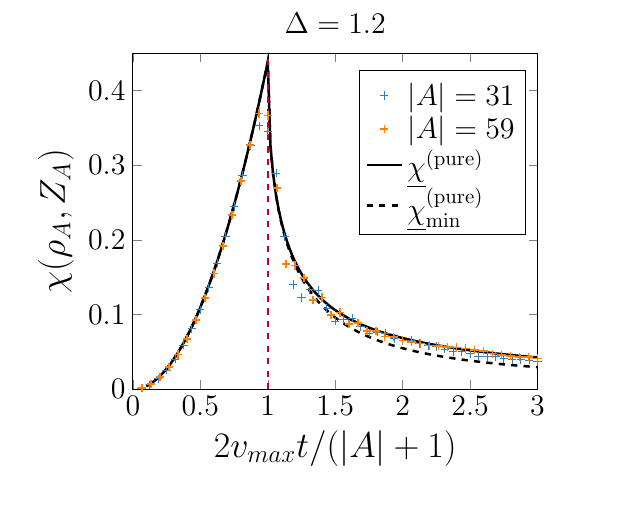}
    \includegraphics[width=0.32\linewidth]{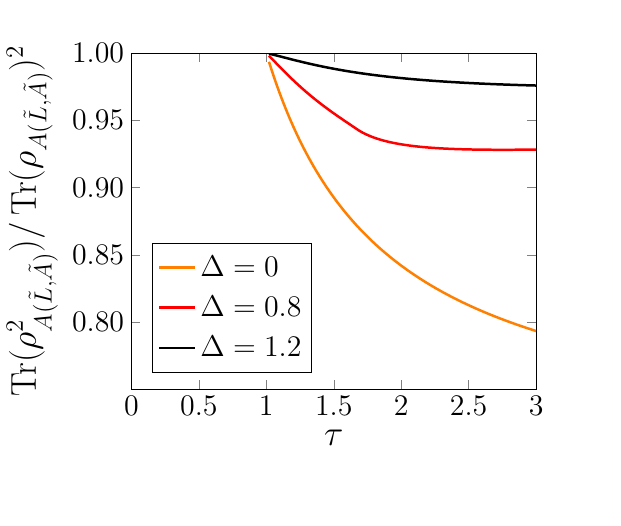}
    \includegraphics[width=0.32\linewidth]{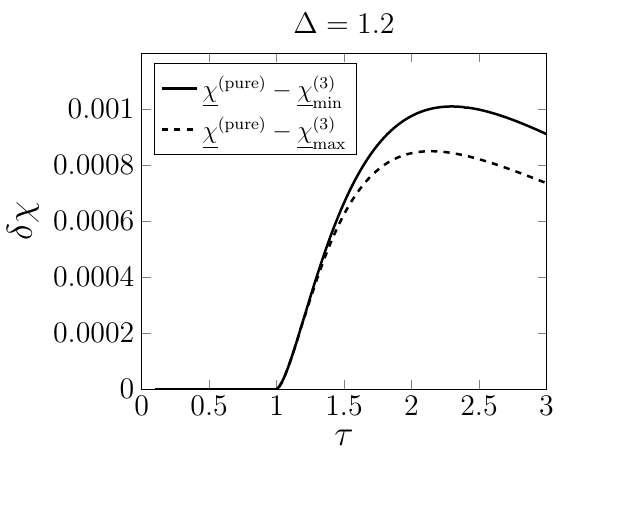}
    \caption{Left: time evolution of the nQFI for two subsystems of lengths $|A| = 31$ and
    $|A|=59$ centered around the initial spin flip, compared with the analytical bounds for the
    nQFI in the scaling limit, see Eq.~\eqref{eq:qfi_pure}. Middle: stationary-phase prediction for the purity of the one particle sector contribution $\rho_{A(\tilde L, \tilde A)}$.
    Right: higher-order corrections to the
    nQFI, see Eq.~\eqref{eq:qfi_cut}, \eqref{eq:moments} and \eqref{eq:qfi_bounds_cut}.}
    \label{fig:qfi}
\end{figure}

\paragraph*{}
In the left panel of Fig.~\ref{fig:qfi}, we compare the prediction of the ``pure approximation"
with exact numerical data, together with the predicted lower bound which is obtained directly from
the eigenvalues of $\rho_{A(\tilde L, \tilde A)}$ using the techniques of section \ref{sec:rdm}.
In the right panel, we exhibit the correction to this approximation due to the mixed nature of the
$(\mathcal L, \mathcal A)$ and $(\mathcal A, \mathcal R)$ sectors by computing the nQFI of the
truncated RDM with $n=3$. Apart from the creation of bound states, the behaviour of the nQFI is
qualitatively the same as in the non-interacting case studied in
Ref~\cite{ferro2025kickingquantumfisherinformation}: if we define the rescaled time $\tau =
2t/|A|$, the nQFI grows like $\tau ^2$ for $\tau \le 1$, then decays as $1/\tau$ at late times as
the particle escapes the subsystem. Crucially, the nQFI remains $\mathcal O(1)$ in the scaling
limit at all intermediate times. This confirms that the local quench generates a macroscopic
quantum superposition of states whose magnetisations differ by values that diverge linearly in time.

\section{Discussion: Link between EA and QFI}
In this paper we presented both the QFI and $U(1)$ EA as asymmetry measures, being able to
quantify coherence between charge sectors. However, they are not equivalent, and we argue that the
relation between them is similar to the one between the variance $\Var(\rho, O)$ of an observable
and its Shannon entropy $H(\rho, O) = -\sum_i p_\rho(O_i) \log p_\rho(O_i)$, where $p_\rho(O_i)$
is the probability of the measurement outcome ``$O = O_i$'' in the state $\rho$. For an extensive
observable $O=\sum_i o_i$ consisting of local involutions $o_i^2 = I$, it is well-known that
\cite{Massey1988Proceedings, cover1999elements}
\begin{equation}\label{eq:entropy_variance}
    H(\rho, O) \le \frac{1}{2} \log \left[2\pi e\left(\frac 1 4 \Var(\rho,
    O)+\frac{1}{12}\right)\right].
\end{equation}
When the state $\rho$ is pure, Ref \cite{mazzoni2025breaking} noted that $H(\rho, O)$ precisely
corresponds to the entanglement asymmetry $\Delta S(\rho, O)$. For a mixed state $\rho = \sum_i
\lambda_i \ket i \bra i$, we can generalise this inequality using the Holevo bound, i.e. the
non-increasing property of the Holevo quantity for the decomposition $\{\lambda_i, \ket i\bra i\}$
under the Completely Positive Trace Preserving twirling operation $\mathcal G_O$
\cite{nielsen2000quantum}:
\begin{equation}
    S \big(\sum_i \lambda_i \mathcal G_O(\ket i \bra i)\big) - \sum_i \lambda_i S\big(\mathcal
    G_O(\ket i \bra i)\big) \le S \big(\sum_i \lambda_i \ket i \bra i \big) - \sum_i \lambda_i S
    \big(\ket i \bra i\big),
\end{equation}
which can be rewritten as
\begin{equation}
    S(\mathcal G_O (\rho))-S(\rho) \le  \sum_i \lambda_i S(\mathcal G_O(\ket i \bra i)) .
\end{equation}
The term on the left-hand side of the inequality is precisely the entanglement asymmetry
associated to the $U(1)$ group generated by $O$. In the right hand side term, since $\ket i$ is
pure the Von-Neumann entropy of $\mathcal{G}_O(\ket i \bra i)$ is the Shannon entropy associated
to $O$ in the state $\ket i$, or equivalently the asymmetry $\Delta S (\ket i \bra i, O)$.
Therefore, one can apply Eq.~\eqref{eq:entropy_variance} which gives

\begin{equation}
    \Delta S(\rho, O) \le \sum_i \lambda_i \frac{1}{2} \log \left[2\pi e(\frac 1 4 \Var(\ket i \bra i,
    O)+\frac{1}{12})\right].
\end{equation}
Using the concavity of the logarithm, we find

\begin{equation}
    \Delta S(\rho, O) \le \frac{1}{2} \log \left[\sum_i \lambda_i \left(2\pi e(\frac 1 4 \Var(\ket i \bra i,
    O)+\frac{1}{12})\right)\right].
\end{equation}
This inequality holds for any decomposition $\{\lambda_i, \ket i\bra i\}$ of the reduced density
matrix, in particular for the one that minimises the mean variance $\sum_i \lambda_i \Var(\ket i
\bra i, O)$. Using now the property that the QFI gives the convex roof of the variance
\cite{Toth2013Extremal, yu2013quantum}, $\frac 1 4 F_Q(\rho, O) = \inf_{\{\lambda_i, \ket
i\}}\sum_i \lambda_i \Var(\ket i \bra i, O)$, we finally obtain that 

\begin{equation}\label{eq:bound_asym_qfi}
    \Delta S(\rho, O) \le \frac{1}{2} \log \left[2\pi e\left(\frac{1}{16} F_Q(\rho,
    O)+\frac{1}{12}\right)\right].
\end{equation}

This inequality could be useful as a lower bound on the QFI. Indeed, we can rewrite it for a
subsystem $A$ as 

\begin{equation}\label{eq:bound_asym_qfi_2}
    \frac 1 4 F_Q(\rho_A, O_A)\ge \frac{2}{\pi e}e^{2\Delta S(\rho_A, O_A)}-\frac 1 3,
\end{equation}
which automatically implies super-linear scaling of the QFI in the settings where $\Delta
S(\rho_A, O_A) \sim \alpha \log ||O_A||$ with $\alpha > \frac 1 2 $, see for instance
Ref.~\cite{mazzoni2025breaking}.
 For an extensive $U(1)$ charge, this scaling is excluded in any state with short-range correlations, since the variance in itself is then at most extensive.
Like the super-extensive QFI, it can only appear in the presence of algebraically decaying correlation or breaking of clustering.
The converse is not true: breaking clustering does not guarantee enhanced EA, and the bound is not always saturated.

In our setting, the usefulness of this bound depends on the probability $p^S-p_0^S(\tau)$ of having at least one scattering domain wall inside the subsystem, see Eq.~\eqref{eq:lim_EA}.
The bound therefore predicts non-zero QFI in the scaling limit whenever $p^S-p_0^S(\tau)>0$, and super-extensive QFI when $p^S-p_0^S>1/2$.
However, it only detects quadratic QFI, i.e.
a non-zero normalised QFI $\chi(\rho_A, O_A) = F_Q(\rho_A, O_A)/4||O_A||^2$, when $p^S-p_0^S(\tau)=1$.
For the single spin-flip protocol considered here, this last condition is only satisfied when $\Delta= 0$ and when $\tau < 1$, i.e. when the subsystem contains the light-cone.

\section{Conclusion}
In this paper, we investigated the effect of interaction on the rich phenomenology of a local
quench on an SSB ground state. We studied this setting by considering the dual XXZ Hamiltonian,
which is exactly solvable by coordinate Bethe Ansatz and is analytically tractable after a local
quench. We first showed that the domain-wall nature of the excitations amplify quantum
interferences, leaving clear imprints of the scattering phase on the macroscopic profiles of the
transverse magnetisation and its correlations. Then, we studied the quantum properties of these
correlations under the lens of asymmetry measures, effectively quantifying the coherence between
charge sectors of the magnetization. The leading behaviour of both the entanglement asymmetry and
QFI can be computed in the limit of large subsystems, showing that subsystems around the
light-cone fall in quantum superpositions of a number of magnetisation sectors diverging linearly
in time, and that these sectors correspond to macroscopically different magnetisations. Finally,
we derived several bounds concerning the EA and QFI, which can be used to greatly simplify the
analytical and numerical cost of accessing these quantities thanks to equivalent ones. Since the
analysis relies on the two-particle Bethe Ansatz, this framework is very general and can in fact
also be applied to non-integrable models, as long as they have kink-like excitations. A natural
follow-up to this study would thus be to investigate the effect of integrability breaking within
this picture, and in particular to see if this framework can be used to access the Entanglement
Asymmetry and Quantum Fisher Information after more general local quenches, that generate a higher
number of quasiparticles excitations.

\section{Acknowledgements}
I thank Maurizio Fagotti for collaboration on the first stages of the project, as well as Luca
Capizzi and Sara Murciano for useful discussions.

\appendix

\section{Thermodynamic limit for transition amplitudes}\label{ap:thermo_amplitudes}
We show here the main steps to compute the transition amplitude $\prescript{}{s}{\bra{f_1
f_2}}e^{-iHt} \ket{l_1 l_2}_{s'}$ in the thermodynamic limit. Since $[H, P]=0$, this amplitude is
non-zero only when the two parity sectors match, $s=s'$. The CBA computation of the amplitude in
each parity sector is the same as in the XXZ model - see \cite{Franchini2017An} for a
comprehensive review - except that i) the elementary excitations are domain walls instead of spin
flips and ii) the Bethe equations are modified by the antiperiodic boundary conditions in the
$P=-1$ sector. 

We start by determining the eigenbasis of $H$ in the $Q=2$ sector. For a spin chain of size $N$
with periodic boundary conditions, we can construct the eigenstates by considering wavefunctions
of the form 
\begin{equation}\label{eq:ansatz}
    \ket{k_1 k_2}_s = Z^{(N)}(k_1, k_2)\sum_{l_1< l_2=1/2}^{N-1/2} C_{l_1 l_2}^{k_1 k_2}\ket{l_1
    l_2}_s, \text{ with }C_{l_1 l_2}^{k_1 k_2}= S(k_1, k_2) e^{i(k_1 l_1 + k_2 l_2)}+e^{i(k_2 l_1
    + k_1 l_2)}.
\end{equation}

Solving the eigenvalue equation $H\ket{k_1, k_2}  =(\epsilon_{k_1} + \epsilon_{k_2})\ket{k_1, k_2}
$ imposes the expression of the bare energy $\epsilon_k = -\cos k$ and of the scattering phase,
\begin{equation}
    S(k_1, k_2) = -\frac{1+e^{i(k_1+k_2)}-2\Delta e^{i k_1}}{1+e^{i(k_1+k_2)}-2\Delta e^{i k_2}}.
\end{equation}
As mentioned in section \ref{sec:CBA}, the boundary conditions depend on the parity sector,
$\ket{l_1 N+\frac 1 2}_\pm = \pm \ket{\frac 1 2 l_1}$. Therefore, so do the Bethe equations which
impose the quantisation of the momenta:
\begin{equation}\label{eq:betheEq}
    e^{ik_1 N} = s S(k_1, k_2) \text{ and } e^{-ik_2 N} = s S(k_1, k_2), 
\end{equation}
where $s=1$ (resp. $-1$) in the sector with (anti-)periodic boundary conditions. 

We now use the states of Eq.~\eqref{eq:ansatz} to write a resolution of the identity in the
two-particle sector. One can see that for every solution $(k_1, k_2)$ of the Bethe equations,
$S(k_1, k_2) = 1/S(k_2, k_1)$ so $\ket{k_1 k_2}_s$ is colinear with $\ket{k_2 k_1}_s$. Therefore,
writing $\mathcal{B}_s^{(N)}$ the set of solutions of the Bethe equations~\eqref{eq:betheEq} the
transition amplitude reads
\begin{equation}\label{eq:amplitude_general}
    \prescript{}{s}{\bra{f_1 f_2}}e^{-iHt} \ket{l_1 l_2}_s = \frac 1 2 \sum_{(k_1, k_2)\in
    \mathcal{B}_s^{(N)}} |Z^{(N)}(k_1, k_2)|^2 C_{f_1 f_2}^{k_1 k_2}(C_{l_1 l_2}^{k_1
    k_2})^*e^{-i(\epsilon_{k_1}+\epsilon_{k_2})t}.
\end{equation}
\subsection{Real momenta}
We first focus on the solutions $(k_1, k_2) \in \mathbb{R}^2$. For these solutions, the scattering
phase has unit modulus, $|S(k_1, k_2)| = 1$, and satisfies $S(k_1, k_2)^{-1} = S(k_1, k_2)^* =
S(k_2, k_1)$. We can thus write the term in Eq~\eqref{eq:amplitude_general} as
\begin{multline}
    C_{f_1 f_2}^{k_1 k_2}(C_{l_1 l_2}^{k_1 k_2})^* \underset{(k_1, k_2)\in \mathbb{R}^2}{=}
    \left(S(k_1, k_2)e^{i(k_1 f_1 + k_2 f_2)}+e^{i(k_2 f_1 + k_1 f_2)}\right) \times
    \left(S^*(k_1, k_2)e^{-i(k_1 l_1 + k_2 l_2)}+e^{-i(k_2 l_1 + k_1 l_2)}\right) \\
    =\left[e^{ik_1(f_1-l_1)}e^{ik_2 (f_2-l_2)}+ e^{ik_2(f_1-l_1)}e^{ik_1
    (f_2-l_2)}\right.\\ \left.+S(k_1, k_2) e^{ik_1(f_1-l_2)}e^{ik_2 (f_2-l_1)}+S(k_2, k_1)
    e^{ik_2(f_1-l_2)}e^{ik_1 (f_2-l_1)}\right].
\end{multline}
These terms are however two by two symmetric under the exchange $k_1 \leftrightarrow k_2$,
therefore 
\begin{equation}\label{eq:sum_reals}
    \frac 1 2 \sum_{(k_1, k_2)\in \mathcal{B}_s^{(N)}\cap \mathbb{R}^2}|Z^{(N)}(k_1, k_2)|^2
    C_{f_1 f_2}^{k_1 k_2}(C_{l_1 l_2}^{k_1 k_2})^*e^{-i(\epsilon_{k_1}+\epsilon_{k_2})t}=
    \sum_{(k_1, k_2)\in \mathcal{B}_s^{(N)}\cap \mathbb{R}^2}|Z^{(N)} (k_1, k_2)|^2
    \mathcal{A}^S_{l_1, l_2\to f_1, f_2}(k_1, k_2, t),
\end{equation}
where
\begin{equation}
    \mathcal{A}_{l_1, l_2\to f_1, f_2}^S(k_1, k_2, t)= (e^{ik_1(f_1-l_1)}e^{ik_2 (f_2-l_2)}+S(k_1,
    k_2) e^{ik_1(f_1-l_2)}e^{ik_2 (f_2-l_1)})e^{-i(\epsilon_{k_1}+\epsilon_{k_2})t}.
\end{equation}
The normalisation factor is given by
\begin{multline}\label{eq:defZ}
    |Z^{(N)} (k_1, k_2)|^{-2}= \sum_{l_1< l_2=1/2}^{N-1/2}C_{l_1 l_2}^{k_1 k_2}(C_{l_1 l_2}^{k_1
    k_2})^*
    \\ = N^2 \left[1+\frac 1 N \frac{1-\cos(k_1-k_2)-\cos(k_2-k_1+\theta_{k_1k_2})+\cos
    \theta_{k_1 k_2}}{\cos(k_1-k_2)-1}+\frac{1}{N^2}\frac{\cos(k_1 N - k_2 N - \theta_{k_1
    k_2})-\cos \theta_{k_1 k_2}}{\cos(k_1-k_2)-1}\right]\\
    = N^2 \left[1+\frac 1 N \frac{1-\cos(k_1-k_2)-\cos(k_2-k_1+\theta_{k_1k_2})+\cos \theta_{k_1
    k_2}}{\cos(k_1-k_2)-1}\right],
\end{multline}
where $e^{i\theta_{k_1 k_2}} = S(k_1, k_2)$ and we imposed the Bethe equations \eqref{eq:betheEq}
to obtain the last line. We take Eq.~\eqref{eq:defZ} to be the definition of $Z^{(N)}(k_1, k_2)$. The
sum~\eqref{eq:sum_reals} can be carried out explicitly exploiting the fact that the total momentum
is conserved. Indeed, making the following change of variables: $\kappa = \frac{1}{2}(k_1+k_2)$
and $\delta = \frac{1}{2}(k_1-k_2)$, the Bethe equations can be rewritten as
\begin{equation}\label{eq:betheEq2}
    (e^{i\kappa N} = s \text{ and } e^{i\delta N} =  S_\kappa(\delta)) \text{ or }(e^{i\kappa N} =
    -s \text{ and } e^{i\delta N} = -S_\kappa(\delta)), 
\end{equation}
where $S_\kappa(\delta) = S(\kappa+\delta, \kappa-\delta)$.
From this new set of Bethe Equations it is clear that $\kappa$ admits the same quantisation as a
free model with either periodic or antiperiodic boundary conditions. The domain of integration
$k_1, k_2\in (-\pi, \pi)$ translates to $\kappa \in (-\pi, \pi)$ and $\delta \in (-\pi+|\kappa|,
\pi-|\kappa|)$. All the difficulty lies in selecting the $\delta$ solutions of
Eq.~\eqref{eq:betheEq2} in that domain, which can be done using a residue theorem. Indeed, 
\begin{multline}
    \sum_{(k_1, k_2)\in \mathcal{B}_s^{(N)}\cap \mathbb{R}^2} |Z^{(N)}(k_1, k_2)|^2 
    \mathcal{A}_{l_1, l_2\to f_1, f_2}^S(k_1, k_2, t) \\= \lim_{\epsilon\to 0} \sum_{n = 1}^N
    \frac{1}{2\pi i}\oint_{\gamma_\epsilon(\kappa_n^+)}|Z^{(N)}(\kappa_n^++\delta,
    \kappa_n^+-\delta)|^2 \frac{siN e^{i\delta N} -  \partial_\delta S_{\kappa_n^+}(\delta)}{s
    e^{i\delta N}-S_{\kappa_n^+}(\delta)} \mathcal{A}_{l_1, l_2\to f_1, f_2}^S(\kappa_n^++\delta,
    \kappa_n^+-\delta, t) d\delta \\ 
    +\lim_{\epsilon\to 0} \sum_{n = 1}^N \frac{1}{2\pi i}\oint_{\gamma_\epsilon(\kappa_n^-)}
    |Z^{(N)}(\kappa_n^-+\delta, \kappa_n^--\delta)|^2 \frac{s iN e^{i\delta N} + \partial_\delta
    S_{\kappa_n^-}(\delta)}{s e^{i\delta N}+S_{\kappa_n^-}(\delta)}  \mathcal{A}_{l_1, l_2\to f_1,
    f_2}^S(\kappa_n^-+\delta, \kappa_n^--\delta, t)  d \delta,
\end{multline}
where $\kappa_n^+ = \frac{2\pi n}{N}$, $\kappa_n^- = \frac{2\pi (n+1/2)}{N}$ and the contour is
given by
\begin{equation}
    \gamma_\epsilon(\kappa) : -\pi+|\kappa|-i\epsilon \rightarrow \pi-|\kappa|-i\epsilon
    \rightarrow \pi-|\kappa|+i\epsilon \rightarrow -\pi+|\kappa|+i\epsilon \rightarrow
    \pi-|\kappa|+i\epsilon.
\end{equation}
Taking the thermodynamic limit $N\to \infty$, the integrand on the interval
$-\pi+|\kappa|+i\epsilon \rightarrow \pi-|\kappa|+i\epsilon$ vanishes as $N\to \infty$ like
$1/N^2$, while on the interval $-\pi+|\kappa|-i\epsilon \rightarrow \pi-|\kappa|-i\epsilon$ it
simplifies to $|Z^{(N)}_{\kappa_n^+}(\delta)|^2 (siN e^{i\delta N} -  \partial_\delta
S_{\kappa_n^+}(\delta))/(s e^{i\delta N}-S_{\kappa_n^+}(\delta)) \to i/N$. Taking the limit $N\to
\infty$ then $\epsilon\to 0$ in the sum gives
\begin{multline}
    \lim_{N\to \infty}\sum_{(k_1, k_2)\in \mathcal{B}_s^{(N)}\cap \mathbb{R}^2}|Z^{(N)}(k_1,
    k_2)|^2 \mathcal{A}_{l_1, l_2\to f_1, f_2}^S(k_1, k_2, t) = \lim_{\epsilon\to 0} \sum_{n =
    1}^N \frac{1}{2\pi N}\int_{-\pi+|\kappa|+i\epsilon }^{\pi-|\kappa|+i\epsilon}d \delta
    \mathcal{A}_{l_1, l_2\to f_1, f_2}^S(\kappa_n^++\delta, \kappa_n^+-\delta, t)\\ 
    +\lim_{\epsilon\to 0} \sum_{n = 1}^N \frac{1}{2\pi N }\int_{-\pi+|\kappa|+i\epsilon
    }^{\pi-|\kappa|+i\epsilon} d\delta \mathcal{A}_{l_1, l_2\to f_1, f_2}^S(\kappa_n^-+\delta,
    \kappa_n^--\delta, t),
\end{multline}
independently of the parity sector. Finally, we can simplify these sums using the Euler-Maclaurin
formula:
\begin{equation}\label{eq:thermo_sum}
    \lim_{N\to +\infty} \sum_{(k_1, k_2)\in \mathcal{B}_s^{(N)}\cap \mathbb{R}^2}|Z^{(N)}(k_1,
    k_2)|^2 \mathcal{A}_{l_1, l_2\to f_1, f_2}^S(k_1, k_2, t) = \lim_{\epsilon \to 0}
    \frac{1}{2\pi^2} \int_{-\pi}^\pi d\kappa \int_{-\pi+|\kappa|+i\epsilon
    }^{\pi-|\kappa|+i\epsilon}d \delta \mathcal{A}_{l_1, l_2\to f_1, f_2}^S(\kappa+\delta,
    \kappa-\delta, t).
\end{equation}
By changing back to the original variables and keeping the $+i\epsilon$ regularisation implicit,
we simply have
\begin{equation}\label{eq:thermo_sum_final}
    \lim_{N\to +\infty} \sum_{(k_1, k_2)\in \mathcal{B}_s^{(N)}\cap \mathbb{R}^2}|Z^{(N)}(k_1,
    k_2)|^2 \mathcal{A}_{l_1, l_2\to f_1, f_2}^S(k_1, k_2, t) = \frac{1}{4\pi^2} \int_{-\pi}^\pi
    dk_1 dk_2\, \mathcal{A}_{l_1, l_2\to f_1, f_2}^S(k_1, k_2, t).
\end{equation}
\subsection{Complex momenta}
We now consider complex solutions $(k_1, k_2)\notin \mathbb{R}^2$. If the momenta have non-zero
imaginary part, the Bethe equations imply that either $S(k_1, k_2)$ or $S(k_2, k_1)$ is
exponentially small in $N$. Consequently, the solutions must have the form 
\begin{equation}
    (k_1, k_2) = (\frac K 2 + i\eta, \frac K 2-i \eta)+ \mathcal O(e^{-N}),
\end{equation}
with $K, \eta \in \mathbb R$. For momenta of this form, the term $C_{f_1 f_2}^{k_1 k_2}(C_{j_1
j_2}^{k_1 k_2})^*$ takes the form
\begin{multline}
   C_{f_1 f_2}^{k_1 k_2}(C_{j_1 j_2}^{k_1 k_2})^* \underset{(k_1, k_2)=(\frac K 2 + i\eta, \frac K
   2-i \eta)}{=} \left(S(k_1, k_2)e^{i(k_1 f_1 + k_2 f_2)}+e^{i(k_2 f_1 + k_1 f_2)}\right) \times
   \left(S(k_1, k_2)e^{i(k_1 j_1 + k_2 j_2)}+e^{i(k_2 j_1 + k_1 j_2)}\right)^* \\
    =e^{iK(\frac{f_1+f_2}{2}-\frac{j_1+j_2}{2})}\left[S\left(\frac K 2 + i\eta, \frac K 2 -
    i\eta\right) e^{\eta(f_2-f_1)}+e^{-\eta(f_2-f_1)}\right]\left[S\left(\frac K 2 + i\eta, \frac
    K 2 - i\eta\right) e^{\eta(j_2-j_1)}+e^{-\eta(j_2-j_1)}\right].
\end{multline}
In the thermodynamic limit, the physical solutions should stay normalisable. Therefore, to avoid
divergence as $f_2-f_1 \to \infty$ they should correspond to $S(K/2+i\eta, K/2-i\eta) = 0$ (fixing
$\eta>0$), which imposes the value of $\eta$ as a function of $K$:
\begin{equation}
    e^{-\eta(K)} = \frac{\cos (K/2)}{\Delta}, \text{ and }\cos K/2 \le \Delta.
\end{equation}
The bound state contributions are thus given by
\begin{equation}
   C_{f_1 f_2}^{k_1 k_2}(C_{j_1 j_2}^{k_1 k_2})^* \underset{(k_1, k_2)=(\frac K 2 + i\eta, \frac K
   2-i \eta)}{=} e^{iK(\frac{f_1+f_2}{2}-\frac{j_1+j_2}{2})}e^{-\eta(f_2-f_1+j_2-j_1)},
\end{equation}
and the normalisation factor is 
\begin{multline}
    |Z(K/2+i\eta, K/2-i\eta)|^{-2} =\sum_{f_1< f_2=1/2}^{N-1/2} |C_{f_1 f_2}^{k_1 k_2}|^2 \\=
    \sum_{f_1< f_2=1/2}^{N-1/2} e^{-2\eta (f_2-f_1)} = \sum_{f_1 =
    1}^Ne^{-2\eta}\frac{1-e^{-2\eta(N-f_1)}}{1-e^{-2\eta}} = \frac{N}{e^{2\eta}-1} + \mathcal{O}(1).
\end{multline}
Finally, using the Euler-Maclaurin formula we find that
\begin{multline}
    \lim_{N\to \infty}\sum_{(k_1, k_2)\in \mathcal{B}_s^{(N)}\backslash \mathbb{R}^2}|Z^{(N)}(k_1,
    k_2)|^2 C_{f_1 f_2}^{k_1 k_2}(C_{j_1 j_2}^{k_1 k_2})^*e^{-i(\epsilon_{k_1}+ \epsilon_{k_2})t}\\ = \frac{1}{2\pi}\int_{-\pi}^{\pi} dK\,
    (e^{2\eta(K)}-1) e^{iK(\frac{f_1+f_2}{2}-\frac{j_1+j_2}{2})}e^{-\eta(K)(f_2-f_1+j_2-j_1)}
    e^{-iE_Kt}\theta(\eta(K)).
\end{multline}

\section{Scaling limits}\label{ap:scaling}
In this appendix, we show how to obtain the scaling limit of $p_{(X, Y)}(t)$
\subsection{Scattering states contribution}
The contribution of the scattering states can be expressed as
\begin{multline}
    p_{(X, Y)}^S(t) = \bra{\psi(t)} \Pi_{X}^{Y} \ket{\psi(t)} = \sum_{\substack{f_1\in X\\f_2\in
    Y\\ f_1<f_2}}|\psi_{f_1 f_2}(t)|^2 \\=\sum_{\substack{f_1\in X\\f_2\in Y\\ f_1<f_2}}
    \int_{-\pi}^\pi \frac{dk_1 dk_2 dp_1 dp_2}{(2\pi)^4} \tilde \psi(k_1, k_2)\tilde
    \psi^*(p_1, p_2)e^{i[(f_1-\frac 1 2)(k_1-p_1)+(f_2+\frac 1 2)(k_2-p_2)-(\epsilon_{k_1}+\epsilon_{k_2}-\epsilon_{p_1}-\epsilon_{p_2})t]}.
\end{multline}
For identical or disjoint intervals, the symmetry $|\tilde \psi(k_1, k_2)|^2 = |\tilde \psi(k_2, k_1)|^2$ allows to write the leading term of the scaling limit as
\begin{equation}
    \bra{\psi(t)} \Pi_{X}^{Y} \ket{\psi(t)} = \sum_{\substack{f_1\in X\\f_2\in
    Y}}|\psi_{f_1 f_2}(t)|^2 -\frac 1 2 \sum_{\substack{f_1, f_2\in X \cap
    Y}}|\psi_{f_1 f_2}(t)|^2 + o(1).
\end{equation}
Consequently, the two
relevant terms can be expressed as $\sum_{f_1\in A, f_2\in B}|\psi_{f_1 f_2}(t)|^2$. If we write
$A = \llbracket a_l, a_r\rrbracket$ and $B = \llbracket b_l, b_r\rrbracket$ (with $A$ on the left of $B$), we can explicitly
carry out the geometric sums:
\begin{multline}
    \sum_{\substack{f_1\in A\\f_2\in B}}|\psi_{f_1 f_2}(t)|^2 = \int_{-\pi}^\pi \frac{dk_1 dk_2
    dp_1 dp_2}{(2\pi)^4}
e^{-i(k_1-p_1)/2+i(k_2-p_2)/2}
\left( \tilde \psi(k_1, k_2)\tilde \psi^*(p_1, p_2)\right.\\
    \left.\frac{e^{i a_l(k_1-p_1)}-e^{i(a_r+1)(k_1-p_1)}}{1-e^{i(k_1-p_1)}}\frac{e^{ib_l(k_2-p_2)}-e^{i(b_r+1)(k_2-p_2)}}{1-e^{i(k_2-p_2)}}e^{-i(\epsilon_{k_1}+\epsilon_{k_2}-\epsilon_{p_1}-\epsilon_{p_2})t}\right).
\end{multline}
The integrand is finite at $k_1=p_1$ and $k_2=p_2$, but diverges in the scaling limit as $|A|,
|B|\to \infty$. Therefore, if we define the new set of variables $\kappa_{1, 2} = (k_{1, 2}+p_{1,
2})/2$ and $\delta_{1, 2} = (k_{1, 2}-p_{1, 2})/2$ we can expand the integrand around $\delta_1,
\delta_2\sim 0$:
\begin{multline}\label{eq:pAB_scat}
    \sum_{\substack{f_1\in A\\f_2\in B}}|\psi_{f_1 f_2}(t)|^2 = \frac{1}{16\pi^4}\int_{-\pi}^\pi
    d\kappa_1 d\kappa_2 |\tilde \psi(\kappa_1, \kappa_2)|^2\\ \times
    \int_{-\pi+|\kappa_1|}^{\pi-|\kappa_1|}d \delta_1  \frac{e^{i2t\delta_1 [a_l/t
    -\epsilon'_{\kappa_1}]}-e^{it2\delta_1[a_r/t
    -\epsilon'_{\kappa_1}]}}{-i\delta_1}\int_{-\pi+|\kappa_2|}^{\pi-|\kappa_2|}d \delta_2
    \frac{e^{i2t\delta_2 [b_l/t -\epsilon'_{\kappa_2}]}-e^{i2t\delta_2[b_r/t
    -\epsilon'_{\kappa_2}]}}{-i\delta_2}+\mathcal O(t^{-1}),
\end{multline}
where we discarded the derivative $\partial_{k_1} S(k_1, k_2)\ll t$ in the scaling limit as a
$\mathcal O(t^{-1})$ correction. The integrands do not diverge at $\delta_1, \delta_2 \to 0$.
Therefore, we can sum the principal values of the individual terms:
\begin{equation}
    \mathcal P\int_{-\pi+|\kappa|}^{\pi-|\kappa|}d \delta  \frac{e^{it\delta x}}{-i\delta} =
    -\int_{-\pi+|\kappa|}^{\pi-|\kappa|}d \delta \frac{\sin(t\delta x)}{\delta}.
\end{equation}
Using now the change of variable $\Lambda = t\delta $, we get
\begin{equation}\label{eq:int_rep_sgn}
     -\int_{-\pi+|\kappa|}^{\pi-|\kappa|}d \delta \frac{\sin(t\delta x)}{\delta} \underset{t\to
     \infty}{=}-\int_{-\infty}^{+\infty}d\Lambda\, \frac{\sin(\Lambda x)}{\Lambda} = -\pi \sgn(x). 
\end{equation}
Inserting Eq.~\eqref{eq:int_rep_sgn} into Eq.~\eqref{eq:pAB_scat} yields
\begin{equation}
     \sum_{\substack{f_1\in A\\f_2\in B}}|\psi_{f_1 f_2}(t)|^2 \to \frac{1}{4\pi^2}\int_{-\pi}^\pi
     d\kappa_1 d\kappa_2 |\tilde \psi(\kappa_1, \kappa_2)|^2
     \theta(v_{k_1}t-a_l)\theta(a_r-v_{k_1}t)\theta(v_{k_2}t-b_l)\theta(b_r-v_{k_2}t).
\end{equation}

\subsection{Bound states contribution}
The contribution of the bound states, whose wavefunction is given by
\begin{equation}
    \nu_{f_1,f_2}(t) = \frac{1}{2\pi}\int_{-\pi}^{\pi} dK\, \tilde \nu_{f_2-f_1}(K)
    e^{iK\frac{f_1+f_2}{2}}e^{-iE_Kt},
\end{equation}
reads
\begin{multline}
    p^B_{(X, Y)}(t) = \bra{\nu(t)} \Pi_{X}^{Y} \ket{\nu(t)} = \sum_{\substack{f_1\in X\\f_2\in
    Y\\ f_1<f_2}}|\nu_{f_1 f_2}(t)|^2 \\=\sum_{\substack{f_1\in X\\f_2\in Y\\ f_1<f_2}}
    \frac{1}{(2\pi)^2} \left(\int_{-\pi}^\pi dK dP \,\tilde \nu_{f_2-f_1}(K) \tilde
    \nu_{f_2-f_1}(P)e^{i[(K-P)(f_1+f_2)/2-({E_K}-{E_P})t]}\right).
\end{multline}
We can again separate the sum between $X\backslash Y$, $Y\backslash X$ and $X\cap Y$. Because of
the exponential decay of $\tilde \nu_{f_2-f_1}(K)$ with $f_2-f_1$, if $f_1\in X\backslash Y$ and
$f_2 \in Y\backslash X$ the contribution goes to zero in the scaling limit. The only non-zero
contribution is given by 
\begin{multline}
    p^B_{(X, Y)}(t) \approx \sum_{\substack{f_1, f_2\in X\cap Y\\ f_1<f_2}} \frac{1}{(2\pi)^2}
    \left(\int_{-\pi}^\pi dK dP \,\tilde \nu_{f_2-f_1}(K) \tilde
    \nu_{f_2-f_1}(P)e^{i[(K-P)(f_1+f_2)/2-(E_K-E_P)t]}\right).
\end{multline}
Up to exponentially small corrections in time, we can write
\begin{multline}
    p^B_{(X, Y)}(t) \approx \sum_{\substack{f_1\in X\cap Y}}\sum_{d=1}^{+\infty}
    \frac{1}{(2\pi)^2} \left(\int_{-\pi}^\pi dK dP \,\tilde \nu_{d}(K) \tilde
    \nu_{d}(P)e^{i[(K-P)(f_1+d/2)-(E_K-E_P)t]}\right)\\
    = \sum_{d=1}^{+\infty} \frac{1}{(2\pi)^2} \left(\int_{-\pi}^\pi dK dP \,\tilde \nu_{d}(K)
    \tilde \nu_{d}(P)\frac{e^{il(K-P)}-e^{i(r+1)(K-P)}}{1-e^{i(K-P)}}e^{i[(K-P)d/2-(E_K-E_P)t]}\right)
\end{multline}
Making the change of variable $K= \tilde K + \delta$, $P = \tilde K - \delta$ we expand around the
main contribution at $\delta \sim 0$:
\begin{equation}
    p^B_{(X, Y)}(t) \approx \sum_{r=1}^{+\infty} \frac{1}{(2\pi)^2} \left(\int_{-\pi}^\pi d\tilde
    K \int d\delta \, |\tilde \nu_{r}(\tilde K)|^2 \frac{e^{2it\delta [l/t -E'_{\tilde
    K}]}-e^{2it\delta[r/t -E'_{\tilde K}]}}{-i\delta} \right).
\end{equation}
Using now that
\begin{equation}
    \sum_{r=1}^{+\infty}|\tilde \nu_{r}(\tilde K)|^2 = (1-e^{-2\eta(K)})\theta(\eta(K)),
\end{equation}
we finally find
\begin{equation}
    \underline{p}^B_{(\mathcal X, \mathcal Y)}= \frac{1}{(2\pi)} \left(\int_{-\pi}^\pi d\tilde
    K\,(1- e^{-2\eta(\tilde K)} )\Theta[E'_{\tilde K} \in \mathcal X\cap \mathcal Y]\theta(\eta(K)) \right).
\end{equation}

\section{Bounds on QFI}\label{ap:bounds}
\subsection{From truncated RDM}
Consider a (normalised) density matrix acting on a Hilbert space of dimension $D$ ($D$ can be
infinite) $\rho=\sum_{j = 1}^{D} \lambda_j \ket j \bra j$ with eigenvalues organised in decreasing
order, $\forall j \,\lambda_j \ge \lambda_{j+1}$. If $\rho$ is rank $R$ with $R<D$, it admits such
an expression where $\lambda_i = 0$ for $i> R$. The QFI of $\rho$ with respect to an observable
$\mathcal O$ is given by 
\begin{equation}\label{eq:qfi_de}
    \frac 1 4 F_Q(\rho, \mathcal O) = \frac 1 2 \sum_{i, j=1}^D \frac{(\lambda_i -
    \lambda_j)^2}{\lambda_i + \lambda_j}|\bra{i}\mathcal O \ket j|^2.
\end{equation}
The term $\frac{(\lambda_i - \lambda_j)^2}{\lambda_i + \lambda_j}$ tends to $0$ when $\lambda_i,
\lambda_j\to 0$ but not when only one of the eigenvalues is set to zero, therefore the double sum
spans the full Hilbert space. Separating zero from non-zero eigenvalues and inserting a resolution
of the identity, we can re-express Eq.\eqref{eq:qfi_de} as a sum over non-zero eigenvalues of the
density matrix:
\begin{equation}\label{eq:qfi_alt}
    \frac 1 4 F_Q(\rho, \mathcal O) = \sum_{i=1}^{R} \lambda_i \bra i\mathcal{O}^2\ket i  -
    2\sum_{i, j=1}^R \frac{\lambda_i \lambda_j}{\lambda_i + \lambda_j}|\bra{i}\mathcal O \ket j|^2.
\end{equation}
Let us now truncate $\rho$ keeping only its $n$ largest eigenvalues, and write $\rho =
(1-\epsilon_n)\rho^{(n)}+ \epsilon_n \delta \rho^{(n)}$, with $(1-\epsilon_n)\rho^{(n)} =
\sum_{i=1}^n \lambda_i \ket i \bra i$ and $\epsilon_n \delta \rho^{(n)} = \sum_{i=n+1}^D \lambda_i
\ket i \bra i$. Because of the convexity properties of the QFI, 
\begin{equation}
    \frac{1}{4} F_Q (\rho, \mathcal O) \le \frac 1 4 (1-\epsilon_n)F_Q(\rho^{(n)}, \mathcal O) +
    \epsilon_n ||\mathcal O||^2,
\end{equation}
where we used the fact that $\frac 1 4 F_Q(\sigma, \mathcal O)$ is bounded by norm squared of
$\mathcal O$, $||\mathcal O||^2$, for any density matrix $\sigma$. 

Let us now construct a lower bound for $F_Q(\rho, \mathcal O)$ from $F_Q(\rho^{(n)}, \mathcal O)$.
We can see from Eq.~\eqref{eq:qfi_alt} that
\begin{equation}
    \frac 1 4 F_Q (\rho, \mathcal O) = \frac 1 4 (1-\epsilon_n) F_Q(\rho^{(n)}, \mathcal O)+\frac
    1 4 \epsilon_n F_Q(\delta \rho^{(n)}, \mathcal O)-4\sum_{i=1}^n\sum_{j=n+1}^R \frac{\lambda_i
    \lambda_j}{\lambda_i +\lambda_j}|\bra i \mathcal O \ket j|^2.
\end{equation}
Using now that $F_Q(\sigma, \mathcal O)\ge 0$, $\lambda_i \lambda_j/(\lambda_i+\lambda_j) \le
\lambda_j$ and $|\bra i \mathcal O \ket j |^2 \le ||\mathcal O||^2$, we finally find that 
\begin{equation}
    \frac 1 4 F_Q (\rho, \mathcal O)\ge \frac 1 4 (1-\epsilon_n) F_Q(\rho^{(n)}, \mathcal O )-4n\,
    \epsilon_n||\mathcal O||^2.
\end{equation}
\subsection{From variance}
When a density matrix is almost pure, it is useful to similarly bound the QFI from the variance of
the state. To that end, let us re-express the QFI as
\begin{equation}
    \frac 1 4 F_Q(\rho, \mathcal O) = \Var(\rho, \mathcal O) + \sum_{i, j=1}^R \lambda_i \lambda_j
    \bra{i}\mathcal O \ket i \bra{j}\mathcal O \ket j - 2\sum_{i, j=1}^R \frac{\lambda_i
    \lambda_j}{\lambda_i + \lambda_j}|\bra{i}\mathcal O \ket j|^2.
\end{equation}
Doing the same procedure as in the previous subsection with $n=1$ (we only keep the largest
eigenvalue), we find that
\begin{multline}
    \frac 1 4 F_Q(\rho, \mathcal O) = \Var(\rho, \mathcal O)+\lambda_1^2 \bra{1}\mathcal O \ket
    1^2-\lambda_1 \bra{1}\mathcal O \ket 1^2 + 2\lambda_1 \bra{1}\mathcal O \ket 1\sum_{ j=2}^R
    \lambda_j  \bra{j}\mathcal O \ket j\\ -4 \sum_{j=2}^R \frac{\lambda_1
    \lambda_j}{\lambda_1+\lambda_j}|\bra{1}\mathcal O \ket j|^2+\sum_{i, j=2}^R \lambda_i
    \lambda_j \bra{i}\mathcal O \ket i \bra{j}\mathcal O \ket j - 2\sum_{i, j=2}^R \frac{\lambda_i
    \lambda_j}{\lambda_i + \lambda_j}|\bra{i}\mathcal O \ket j|^2.
\end{multline}
Using now again the inequality  $\lambda_1 \lambda_j/(\lambda_1+\lambda_j) \le \lambda_j$, and
$2\sum_{i, j=2}^R \frac{\lambda_i \lambda_j}{\lambda_i + \lambda_j}|\bra{i}\mathcal O \ket j|^2\le
\epsilon_1\Tr(\delta \rho^{(1)}\mathcal O^2)\le \epsilon_1||O||^2$, we get 
\begin{equation}
    \frac 1 4 F_Q(\rho, \mathcal O) \ge \Var(\rho, \mathcal O) +
    (\epsilon_1^2-\epsilon_1-2(1-\epsilon_1)\epsilon_1-4\epsilon_1-\epsilon_1)||O||^2,
\end{equation}
or the simpler bound
\begin{equation}
    \frac 1 4 F_Q(\rho, \mathcal O) \ge \Var (\rho, \mathcal O) - 8\epsilon_1 ||\mathcal O||^2.
\end{equation}
This bound presents the advantage of not requiring any spectral analysis of the density matrix,
since also $\epsilon_1$ can be bounded by the purity of the density matrix: $\epsilon_1 \le 1/2
(1-\sqrt{2\Tr \rho^2-1})$ when $\Tr \rho^2 \ge 1/2$. Therefore, we finally have
\begin{equation}
    \frac 1 4 F_Q(\rho, \mathcal O) \ge \Var (\rho, \mathcal O) - 4\left(1-\sqrt{2\Tr
    \rho^2-1}\right) ||\mathcal O||^2.
\end{equation}

\section{Fourier coefficients of the scattering phase in XXZ model}\label{ap:S_diag}
Since $S(k_1, k_2)$ is a bounded function appearing only as a kernel on a compact domain (i.e.
$[-\pi, \pi]^2$), one can represent it in Fourier basis:
\begin{equation}
    S(k_1, k_2) \equiv \sum_{nm\in \mathbb{Z}} S_{nm} e^{i k_1 n} e^{i k_2 m}
\end{equation}
Let us recall the explicit expression of the scattering phase and recast it as
\begin{equation}
    S(k_1, k_2) = -e^{i(k_1-k_2)}\frac{e^{-ik_1}+e^{ik_2} - 2\Delta}{e^{ik_1}+e^{-ik_2} - 2\Delta}
\end{equation}
 
\subsection{$\Delta > 1$:}
When $\Delta> 1$, $|e^{ik_1}+e^{-ik_2}|< 2\Delta $ and one can easily obtain the Fourier
coefficients by expanding 
\begin{equation}
    S(k_1, k_2) = -e^{i(k_1-k_2)}\frac{e^{-ik_1}+e^{ik_2} - 2\Delta}{e^{ik_1}+e^{-ik_2} - 2\Delta}
    = \frac{e^{i(k_1-k_2)}}{2\Delta}(e^{-i k_1}+ e^{i
    k_2}-2\Delta)\sum_{j=0}^{+\infty}\left(\frac{e^{i k_1}+e^{-ik_2}}{2\Delta}\right)^j, 
\end{equation}
which gives 
\begin{equation}
    S_{jp} =\begin{cases}
        \frac{
\left( j^2 - j + p + (4\Delta^{2} - 2)jp + p^{2} \right)
}{
\left( j^2-j + p - 2jp + p^{2} \right)
}(2\Delta)^{p - j}\dbinom{j - p}{j}\text{ if } j>0 \text{ and } p<0\\
 (2\Delta)^p \text{ if } (j=0 \text{ and } p<0) \text{ or }  (2\Delta)^{-j} \text{ if } (p=0 \text{ and } j>0) \\
        0 \text{ otherwise}
    \end{cases}  
\end{equation}
\subsection{$\Delta<1$}
When $\Delta<1$, the most convenient way to obtain the Fourier coefficients of the scattering
phase is to compute the integral 
\begin{equation}
    S_{nm} = \frac{1}{4\pi^2}\int_{-\pi}^\pi dk_1 dk_2 S(k_1, k_2)e^{-i(k_1 n + k_2 m)}
\end{equation}
using residue theorem. First of all, notice that the imaginary part of the integrand is odd for
$k_1, k_2\in \mathbb R$. Therefore, the Fourier coefficients are real. Let us now make the change
of variable $\kappa=\frac{k_1 + k_2}{2}$ and $\delta = \frac{k_1 - k_2}{2}$. We have
\begin{equation}\label{eq:fourier_scat_2}
    S_{nm} = \frac{1}{\pi^2}\int_{0}^\pi d\kappa \int_{-\pi+\kappa}^{\pi-\kappa} d\delta \, \Re
    \left( S_\kappa(\delta) e^{-i \kappa ( n + m)}e^{-i\delta(n-m)}\right).
\end{equation}
where 
\begin{equation}
    S_\kappa(\delta)  =S(\kappa+\delta, \kappa-\delta) = -\frac{\cos \kappa - \Delta
    e^{i\delta}}{\cos \kappa - \Delta e^{-i\delta}}.
\end{equation}
We can now evaluate \eqref{eq:fourier_scat_2} by deforming the contour of integration of $\delta$
using residue theorem on 
\begin{equation}
 \Gamma: -\pi+\kappa\to \pi-\kappa\to \pi-\kappa+i\Lambda \to
-\pi+\kappa+i\Lambda\to -\pi+\kappa \to -\pi+\kappa,
\end{equation}
with $\Lambda\in \mathbb{R}$. The scattering phase satisfies
\begin{equation}\label{eq:scattering_symmetry_1}
    S_\kappa(\pi-\kappa+i\eta)e^{-i(\pi-\kappa+i\eta)(n-m)} =
    (S_\kappa(-\pi+\kappa+i\eta)e^{-i(-\pi+\kappa+i\eta)(n-m)})^*.
\end{equation}

    Therefore, we can write the vertical contributions of the contour integral as
\begin{multline}
V_\kappa=
\int_{\pi-\kappa}^{\pi-\kappa+i\Lambda}
d\delta\,S_\kappa(\delta)e^{-i\delta(n-m)}-
\int_{-\pi+\kappa}^{-\pi+\kappa+i\Lambda}d\delta\,S_\kappa(\delta)e^{-i\delta(n-m)}
\\
= \int_0^\Lambda d\eta \left[i S_\kappa(\pi-\kappa + i\eta) e^{-i(\pi-\kappa+i\eta)(n-m)} + c.c\right].
\end{multline}
$d\delta$ is imaginary, therefore $V_\kappa$ is real.
Furthermore, using that $S_{\pi-\kappa}(\delta)=S_\kappa(\delta+\pi)$, $V_\kappa$ must satisfy $V_{\pi-\kappa} = -(-1)^{n-m} V_\kappa$.
The vertical contribution appearing in the Fourier coefficient $S_{nm}$ is therefore
\[
\operatorname{Re}\left[e^{-i\kappa(n+m)}V_\kappa\right]
=
\cos[(n+m)\kappa]V_\kappa.
\]
Now, the cosine term has the same symmetry: $\cos[(n+m)(\pi-\kappa)]=(-1)^{n-m}\cos[(n+m)\kappa]$.
Therefore, the integrand $\cos((n+m)\kappa) V_\kappa$, which is antisymmetric around $\kappa=\pi/2$, vanishes upon integration:

\begin{equation}
    \int_{0}^{\pi} d\kappa \, \cos((n+m)\kappa) V_\kappa = 0.
\end{equation}

This allows to rewrite \eqref{eq:fourier_scat_2} as

\begin{equation}
    S_{nm} = \Re \frac{1}{\pi^2}\int_{0}^\pi d\kappa\, e^{-i \kappa ( n +
    m)}\left[\int_{-\pi+\kappa+i\Lambda}^{\pi-\kappa+i\Lambda} d\delta \,
    S_\kappa(\delta) e^{-i\delta(n-m)}+ 2\pi i\sgn \Lambda \, R_\Lambda(\kappa)\right],
\end{equation}
where $R_\Lambda(\kappa)$ is the sum of the residues of $S_\kappa(\delta)
e^{-i\delta(n-m)}$ in the contour $\Gamma$. The poles are located at $\delta_\kappa
=i\ln(|\cos\kappa/\Delta|)+\theta(-\cos \kappa/\Delta)\pi $ (modulo $2\pi)$. Therefore,
\begin{equation}
    R_\Lambda(\kappa) = -i\, \theta(\frac{\cos \kappa}{\Delta}) \theta(|\Lambda|-|\Im
    \delta_\kappa|)\theta[\sgn(\Lambda) \sgn (\Im \delta_\kappa)]\left[\left(\frac{\cos
    \kappa}{\Delta}\right)^{n-m-2}-\left(\frac{\cos \kappa}{\Delta}\right)^{n-m}\right].
\end{equation}
Now, when $|\Lambda| \to \infty$, the integrand behaves as $e^{\Lambda (n-m-1)}$.
Therefore, one can cancel the double integral term by taking $\Lambda = +\infty$ when $n-m<1$ or
$\Lambda = -\infty$ when $n-m>1$. 
Either way, we can integrate $ R_\Lambda(\kappa)$ and finally
express the scattering phase in the form
\begin{equation}
    S(k_1, k_2) = \sum_{l_1
    l_2}(\lambda_{l_1+l_2}^{l_2-l_1+2}-\lambda_{l_1+l_2}^{l_2-l_1}-\delta_{l_10}\delta_{l_20})e^{ik_1 l_1}e^{i k_2 l_2}, 
\end{equation}
where 
\begin{equation}
    \lambda_{a}^{b} = \frac{2\sgn (b-\frac 1 2)}{\pi}\int_{0}^{\pi/2}d\kappa\, \left(\frac{\Delta}{\cos
    \kappa}\right)^b \cos(a \kappa)\theta[\sgn(b-\frac 1 2)(\cos \kappa-\Delta)].
\end{equation}
The $n-m = 1$ case require special treatment, but leads to the same final expression.

\bibliography{references.bib}

\end{document}